\newcommand{\vsys}{V_{\rm sys}}

\newcommand{\vlos}{V_{\rm LOS}}

\newcommand{\vrot}{V_{\rm rot}}

\newcommand{\slos}{\sigma_{\rm LOS}}

\newcommand{\sslos}{\sigma_{\ast}^{\rm LOS}}

\newcommand{\sddisk}{\Sigma_{\rm dyn}}

\newcommand{\mldyndisk}{\Upsilon_{\rm dyn}^{\rm disk}}

\newcommand{\mldynkdisk}{\Upsilon_{{\rm dyn},K}^{\rm disk}}
\newcommand{\mls}{\Upsilon_\ast}
\newcommand{\mlsdisk}{\Upsilon_\ast^{\rm disk}}

\newcommand{\mlsk}{\Upsilon_{\ast,K}}

\newcommand{\Fdisk}{\mathcal{F}_{\ast,{\rm max}}^{\rm disk}}

\documentclass[12pt,preprint]{aastex}

\slugcomment{To appear in ApJ}

\begin{document}

\title{The DiskMass Survey. I. Overview}

\author{
Matthew A. Bershady,\altaffilmark{1}
Marc A. W. Verheijen,\altaffilmark{2} 
Rob A. Swaters,\altaffilmark{3} 
David R. Andersen,\altaffilmark{4}
Kyle B. Westfall,\altaffilmark{1,2,5} and
Thomas Martinsson\altaffilmark{2} 
}

\altaffiltext{1}{University of Wisconsin, Department of Astronomy, 475
N. Charter St., Madison, WI 53706; mab@astro.wisc.edu}

\altaffiltext{2}{University of Groningen, Kapteyn Astronomical
  Institute, Landleven 12, 9747 AD Groningen, Netherlands}

\altaffiltext{3}{University of Maryland, Dept. of Astronomy, College
  Park, MD 20742}

\altaffiltext{4}{NRC Herzberg Institute of Astrophysics, 5071 W
Saanich Road, Victoria, BC V9E 2E7}

\altaffiltext{5}{National Science Foundation (USA) International Research Fellow}

\smallskip
\begin{abstract}
  We present a survey of the mass surface-density of spiral disks,
  motivated by outstanding uncertainties in rotation-curve
  decompositions. Our method exploits integral-field spectroscopy to
  measure stellar and gas kinematics in nearly face-on galaxies
  sampled at 515, 660, and 860 nm, using the custom-built SparsePak
  and PPak instruments. A two-tiered sample, selected from the UGC,
  includes 146 nearly face-on galaxies, with $B<14.7$ and disk
  scale-lengths between 10 and 20 arcsec, for which we have obtained
  H$\alpha$ velocity-fields; and a representative 46-galaxy subset for
  which we have obtained stellar velocities and velocity
  dispersions. Based on re-calibration of extant photometric and
  spectroscopic data, we show these galaxies span factors of 100 in
  $L_K$ ($0.03<L/L_K^*<3$), 8 in $L_B/L_K$, 10 in R-band disk central
  surface-brightness, with distances between 15 and 200 Mpc. The
  survey is augmented by 4-70 $\micron$ Spitzer IRAC and MIPS
  photometry, ground-based $UBVRIJHK$ photometry, and \ion{H}{1}
  aperture-synthesis imaging. We outline the spectroscopic analysis
  protocol for deriving precise and accurate line-of-sight stellar
  velocity dispersions.  Our key measurement is the dynamical
  disk-mass surface-density. Star-formation rates and kinematic and
  photometric regularity of galaxy disks are also central products of
  the study. The survey is designed to yield random and systematic
  errors small enough (i) to confirm or disprove the maximum-disk
  hypothesis for intermediate-type disk galaxies, (ii) to provide an
  absolute calibration of the stellar mass-to-light ratio well below
  uncertainties in present-day stellar-population synthesis models,
  and (iii) to make significant progress in defining the shape of dark
  halos in the inner regions of disk galaxies.
\end{abstract}


\section{INTRODUCTION}

A major roadblock in testing galaxy-formation models is the disk-halo
{\it degeneracy}: density profiles of dark matter halos, as inferred
from rotation-curve decompositions, depend critically on the adopted
mass-to-light ratio ($M/L$ or $\Upsilon$) of the disk component. While
the disk mass contribution from atomic gas can be reliably inferred
from 21-cm observations of \ion{H}{1}, molecular gas mass estimates
are less well-determined (Casoli et al. 1998) and more controversial
(Pfenniger \& Combes 1994), and estimating the stellar mass-to-light
ratio ($\mls$) from stellar population synthesis (SPS) models
has long been known to require many significant assumptions (Larson \&
Tinsely 1978). These assumptions include the detailed star-formation
and chemical enrichment history, the initial mass function (IMF), and
accurate accounting of late phases of stellar evolution (e.g., TP-AGB;
Maraston 2005, Conroy et al. 2009).  SPS model estimates of $\mls$ are
still very uncertain for all of these reasons. Despite the
uncertainty, the models are used prevalently in the literature for
estimating stellar masses of nearby and distant galaxies -- even for
systems which are photometrically dominated by quite young stellar
populations, for which it is now well quantified that $\mls$
uncertainties are the greatest (Zibetti et al. 2009). Accurate $\mls$
values are critical for inferring the dark-halo profiles of galaxies
and for tracing the cosmic history of the stellar baryon
fraction. Hence, establishing a zero-point for $\mls$ is central to
gaining an understanding of galaxy structure, formation, and
evolution.

An often-used refuge to circumvent the disk-halo degeneracy in
rotation-curve decompositions is the adoption of the maximum-disk
hypothesis (Van Albada et al. 1985, 1986). Namely, one adopts a
$\mls$ value (typically constant with radius) that maximizes the
contribution to the rotation curve by the disk without exceeding the
observed rotation-curve at any radius. Given the observed
rotation-curve and light-profile shapes of spirals, this results in a
maximum disk contribution to the rotation speed, typically at two
radial disk scale-lengths, of about 85 to 90\% (assuming the
dark-matter halo does not have a hollow core). However, this
hypothesis remains unproven, and there have been suggestions to the
contrary based on a lack of surface-brightness dependence to the
Tully-Fisher relation (TF; Tully \& Fisher 1977) for normal barred and
un-barred spirals (Courteau \& Rix 1999; Courteau et al. 2003) as well
as for low surface-brightness spirals (Zwaan et
al. 1995). Unfortunately these arguments only provide {\it relative}
values of $\mls$ which cannot definitively break the
degeneracy. Indeed, van Albada et al. (1986) used the mere existence
of the TF relation and the apparent disk-halo conspiracy\footnote{This
  is not the same as the disk-halo degeneracy; van Albada et
  al. (1986) refer to the apparent fine-tuning of the distributions of
  luminous and dark matter in galaxies such that the inner rotation
  curve is dominated by luminous matter, but the outer rotation curve
  remains flat.}  to argue in favor of disk maximality. Without an
independent measurement of $\mls$, it is not possible to
determine the structural properties of dark matter halos from
rotation-curve decompositions.

This first paper in a series presents an overview of the DiskMass
Survey (DMS), an effort to make a direct, and absolute kinematic
measurement of the mass surface-density of galaxy disks, calibrate
$\mls$, and determine the shapes of dark matter halos.  In the next
section (\S 2) we illustrate the uncertainties in rotation-curve
decompositions (\S2.1) and SPS model constraints on $\mls$ (\S2.2),
and hence motivate the need for the measurements made in this
survey. The survey methods are described in a historical context in \S
2.3, and then in its modern form (\S 3). In \S 4 we detail the
strategy which meets our survey requirements. In \S 5, selection and
properties of the sample are defined.  The specific observations
undertaken to measure total mass, gas mass, and star-formation rates
and disk-mass surface-density are summarized in \S 6. Key elements of
our methodological framework are outlined in \S 7. These developments
are recapped briefly in \S 8. In an accompanying paper (II) we
establish our expected error budget for the primary survey results:
the mass surface-density of the disk ($\sddisk$), the mass-to-light
ratio of the stellar disk ($\mlsdisk$), and disk mass-fraction,
($\Fdisk$); and in a forthcoming paper we present results and detailed
techniques used to derive the stellar velocity dispersion and mass
decomposition from pilot observations of UGC 6918 (Paper III). The
same galaxy is used here to illustrate central features of our
analysis. All distant-dependent quantities are scaled to H$_0$ = 73 km
s$^{-1}$ Mpc$^{-1}$, after corrections for a flow model described in
\S 5; magnitudes are quoted in the Johnson system or otherwise
referenced to Vega.

\section{Survey Motivation}

\subsection{The Disk-Halo Degeneracy}

As a specific example of the fundamental impasse with existing
rotation curve decompositions, we illustrate in Figure 1 a traditional
mass decomposition of NGC 3198 based on a tilted-ring derived
\ion{H}{1} rotation curve (Begeman 1989), and a
photometrically-calibrated image that samples the stellar light
distribution (here, a $K$ band image from 2MASS; Skrutskie et
al. 2006), much the same as the original work by van Albada et
al. (1985). We have carried out our own elliptical-aperture photometry
on the 2MASS image. To focus discussion, we assume that the rotation
curve is accurate in terms of sampling the projected major axis, and
in being well-corrected for beam-smearing, inclination, and the
possible presence of a warp; similarly, the photometric calibration
also is assumed to be free of systematic error. In the case of NGC
3198 these are excellent assumptions, which is why the galaxy was
chosen, but in general these are separate observational issues that
cannot be ignored.

To make the decomposition with the data as posed, however, we must
still assume the mass surface-density and thickness of the disk, and
either a functional form and shape for the halo density profile (as we
have done here), or adopt a non-parametric inversion solution. There
are four ingredients in the disk-mass surface-density, typical of
mass-decomposition methods: (i) We use the light profile to derive the
stellar mass, assuming a value of $\mls$ and its radial
dependence. Here we assume $\mls$ is constant with radius.  The disk
is given a typical oblateness (e.g., de Grijs \& van der Kruit,
1997). (ii) We use 21-cm aperture-synthesis observations to determine
the atomic gas mass, scaling the \ion{H}{1} mass surface-density by a
factor of 1.4 to account for Helium and metals. (iii) We estimate the
molecular gas mass to be close to 20\% of the \ion{H}{1} mass, based
on CO measurements from Braine et al. (1993).  The spatial
distribution is uncertain, although likely more concentrated than the
\ion{H}{1}. Because of the uncertainties in the conversion and
distribution, we ignore the relatively minor molecular gas
contribution. By this intentional oversight we can only underestimate
the uncertainties in the stellar disk mass. (iv) We ignore the
possible presence of disk dark-matter; it is implicitly subsumed into
$\mls$ or the inferred halo profile. Out of these four points, the
outstanding ingredient is $\mls$.

Figure 1 shows that comparable fits to the observed rotation curve can
be obtained for a factor of $\sim$20 in $\mls$ (a factor of 4.5 in
stellar-disk rotation velocity), while also providing a plausible
range of dark-matter halo profiles. An acceptable fit is obtained even
when setting $\mls = 0$ for the entire disk, as shown by van Albada et
al. (1985). Even though $\mls = 0$ is not astrophysically plausible,
this emphasizes their point that rotation curves can be used to set
{\it upper} limits on $\mls$, but {\it not lower} limits. While we
have ignored the likely possibility that $\mls$ varies with radius,
relaxing our unphysical assumption (i) only improves the quality of
the fits. A formal statistical interpretation of the $\chi^2$
distribution indicates $\mls$ in the $K$ band is constrained to within
factors of 1.8, 2.3, and 3.2 at the 65, 95, and 99\% confidence levels
(CL) for three degrees of freedom (i.e., $\mls$, the halo central
density, and characteristic radial scale). The same spirit of
interpretation also leads to the conclusion that a pseudo-isothermal
density profile is preferred at 95\% CL over the NFW profile (Navarro,
Frenk \& White 1997), the latter motivated by simulations of
structure-formation in a $\Lambda$CDM universe. However, these
simplistic statistical interpretations are invalid because the errors
are non-Gaussian and only describe estimated random
uncertainties. Even this quiescent spiral galaxy has non-axisymmetric
motions contributing to systematic variations in the measured
tangential speed of order, or larger than, the estimated random errors
(see Figure 6 in Begeman 1989).  Until non-axisymmetric motions can be
understood and modeled at the level of 5 km s$^{-1}$ on small physical
scales associated with spiral arms and \ion{H}{2} regions (the best
work in the literature indicates this level of accuracy is not yet
obtainable, e.g., Spekkens \& Sellwood 2007), rotation-curve
decomposition constraints on halo parameters and $\mls$ are weak at
best.

\subsection{Uncertainties in $\mls$ from SPS Models}

Can such a large range in $\mls$ values be accommodated given
our knowledge of stellar evolution?  Historically, we have expected
variations are minimized at longer wavelengths where sensitivity to
the presence of hot, massive stars is reduced. These stars are
luminous, short-lived and span a wide range of stellar $M/L$. Based on
this argument the $K$ band might appear ideal given the observational
challenges of the thermal infrared and the presence of dust emission
long-ward of 5 $\mu$m.  However, Portinari et al. (2004) have advocated
the $I$ band is best for estimating $\mls$ because at 0.8$\mu$m
one diminishes uncertainties due to late-phases of stellar evolution
(e.g., the TP-AGB), while minimizing dependence on star-formation
rates and dust. Bell \& de Jong (2001) showed significant improvements
could be made by considering simple correlations of $\mls$
against color -- both in the optical and near-infrared -- for a wide
range of SPS models.  An exquisite refinement along both of these
lines has been to take a multi-color approach to estimating
$\mls$, with a choice of band-passes (e.g., $g,i,H$) minimizing
the impact of nebular emission, extinction, and TP-AGB (Zibetti et
al. 2009).  Assuming a single IMF, Zibetti et al. find a variance of
0.1 dex in $\mls$ is typical over a wide range of multi-colors,
but with up to 0.2 dex variance for blue colors typical of strongly
star-forming galaxies. Unfortunately, they also find $\mls$ up
to 2.5 times smaller than earlier work (e.g., Bell et al. 2003). This
requires some explanation.  Consistent with the analysis of Bruzual
(2007) and Conroy et al. (2009), they suggest the culprit is a change
in the stellar evolutionary tracks for TP-AGB stars, which we now
believe are longer-lived.

Figure 2 shows what one might expect in the $K$ band for a range of
star-formation histories, formation epochs, metallicity ($0.008 < Z <
0.02$), extinction, enrichment models, and IMF, as adopted from the
ground-breaking work of Bell \& de Jong (2001). These use older SPS
models, yet our illustration serves to show their full range of
astrophysical parameters and several different SPS codes. All of these
cases are astrophysically plausible, and produce colors reasonably
matching those of today's spirals.  As they note, there is a
correlation of $\mlsk$ (the stellar mass-to-light ratio in
the $K$ band) with color, here shown as the long-baseline
optical-near-infrared $B-K$. While this band-pass choice may not be
ideal for the reasons given above, it does serve as an excellent
diagnostic of stellar-populations (e.g., Bershady 1995) and reddening.

For comparison, trends for a simple stellar population (SSP) and
constant star-formation rate (cSFR) with Salpeter IMF (0.1 to 125
M$_\odot$) are also shown in the top panels of Figure 2 as a function
of age from 10$^7$ to 10$^{10}$ years (Bruzual \& Charlot 2003). For
cSFR, $\Upsilon_K$ is also calculated including gas mass in future
star-formation up to 10$^{10}$ years, i.e., here $\Upsilon_K$ is the
mass-to-light ratio of all the baryons in the disk, assuming a disk is
formed with all its baryons from the beginning with no subsequent
accretion or loss. From these trends we may surmise that stellar
populations with characteristic ages below a few giga-years have rapid
changes in $\mlsk$ with age, and potentially extreme differences in
$\Upsilon$ of the disk depending on gas content. This serves as a
reminder that for predicting the baryonic mass of disks, we need a
complete picture of the gas-accretion and re-processing history as
well as a reliable zero-point for $\mls$.

Focusing here on the $\mls$ zero-point, Bell \& de Jong (2001)
state there is a factor of 2 range in $\mlsk$ in today's disks given a
`plausible range' in model parameters. However their Figures 1, 3, and
4 clearly show a wider span over their full parameter inventory.
Based on their compilation we find an 0.5 dex range (a factor of 3) in
$\mlsk$ for a given $B-K$ at the extrema; and 1.2 dex (a factor of 15)
range over $2<B-K<4$, typical of spiral galaxies (e.g., Bershady
1995), again at the extrema.  To further complicate matters, the later
work of Bell et al. (2003) shows significant differences compared to
Bell \& de Jong (2001) both in trend, zero-point and scatter in
$\mlsk$ when calculated using different SPS models, different ranges
of age and metallicity, but the same so-called ``diet'' Salpeter IMF.
Certainly differences in stellar-population mean age is a critical
factor, as seen in our Figure 2. Albeit our estimates include their
full suite of models it is reasonable to ask: What, {\it a priori},
defines a {\it plausible} range or distribution of models given our
extant knowledge of galaxy formation and evolution and its range of
variation? This is a particularly hard question to answer because of
the degeneracy in galaxy colors for a wide range of model parameters.

Portinari et al. (2004) present arguably the most sophisticated
analysis of SPS models applied to estimating $\mls$ from galaxy
colors, including a self-consistent treatment of chemical enrichment.
They are able to break out the different physical effects driving
variations in the $\mls$ model predictions, finding a factor of
2.5 range in $\mls$ for variations in star-formation histories
that yield comparable colors, but only about 25-50\% variation in
$\mls$ from metallicity effects due to different chemical
evolution histories.  However, they find an additional factor of 2
uncertainty in the $\mls$ zero-point from a variety of plausible
IMF. To all of this, recall there is an additional factor of 2 to 2.5
uncertainty in $\mls$ in the near-infrared due to uncertainties
in the lifetime of TP-AGB stars. Combining these effects as random
processes yields a factor of $\sim$4 uncertainty in predicting
$\mls$ from photometry, and as much as a factor of 10 to 20
uncertainty if systematic.  Consequently the path to accurate and
precise mass-decompositions and study of dark-matter halo profiles is
blocked based on a purely photometric approach to disk-mass
estimation.

In this context it is worth noting that Bell \& De Jong (2001), and
hence Bell et al. (2003), calibrate the $\mls$ zero-point by tuning
the low-mass end of their ``diet'' Salpeter IMF to yield disks in the
Verheijen (2001) rotation-curve sample that are close to maximal.  Not
only is the tuned, critical mass-range of the IMF unconstrained for
galaxies outside of the Milky Way, but Conroy et al. (2009) point out
even the Milky Way's IMF determinations have considerable uncertainty
between 0.8-2 M$_\odot$. In this range, the estimation requires
uncertain stellar evolution corrections for solar-neighborhood field
stars.  This now brings us back full circle to the disk-halo
degeneracy and the unproven, maximum-disk hypothesis.

Finally, we would like to know what level of precision and accuracy on
$\mls$ is needed to infer meaningful information about the dark
matter halo. One basic fact to establish is whether a dark halo is
needed at all in the inner regions of spiral galaxies. A closely
related question is whether we can differentiate between a maximum
disk and modified gravity based on measurements in these inner
regions.  Since maximum disks contribute 85-90\% of the rotation at
two disk scale-lengths, it is therefore not surprising that estimates
of the required $\mls$ with modified Newtonian gravity (MOND,
e.g., Sanders \& Verheijen 1998, McGaugh 2005) nearly agree with SPS
models calibrated in this way for normal surface-brightness disks,
particularly given the scatter due to other SPS parameters (see top
panels of Figure 2). To falsify MOND via this line of argument, and
simultaneously to make serious headway into understanding the shape of
dark-matter halo cores, requires a level of precision on the mass
surface-density of the disk, and hence $\mls$, of 30\% or
better, i.e., 10 to 15\% in rotation velocity.

\subsection{Breaking the Disk-Halo Degeneracy}

The path around the decomposition impasse is to make direct, dynamical
estimates of the mass of galaxy disks. One direct and absolute
measurement of the dynamical mass-to-light ratio of a galaxy disk
($\mldyndisk$) can be derived from the vertical component, $\sigma_z$,
of the disk stellar velocity dispersion (van der Kruit \& Searle 1981;
Bahcall \& Casertano 1984) and the vertical thickness of the disk.
Other promising approaches include lensing (e.g., Maller et al. 2000),
and using hydrodynamical modeling of non-axisymmetric gas-flows around
bars (Weiner, Sellwood \& Williams 2001) or spiral arms (Kranz, Slyz
\& Rix 2001). Here we focus on collisionless tracers of the disk
potential. For a locally isothermal disk, where the vertical density
distribution decreases with height $z$ above the mid-plan as ${\rm
  sech}^2(z/z_0)$, $\sigma_z = \sqrt{\pi \, {\rm G} \, \mldyndisk \, I
  \, z_0}$, with $I$ the surface-intensity (flux per unit area), and
$z_0$ the disk vertical scale parameter (van der Kruit \& Searle
1981). Disk-mass surface-density is then
\begin{equation}
\sddisk = \sigma_z^2 / \pi {\rm G} z_0. 
\end{equation}
Other functional forms for the vertical distribution modify these
relations by a small change in scale factor.\footnote{We consider the
  more general case in Paper II.} The observational conundrum is how
to determine a parameter like $z_0$, best seen at edge-on projections,
and $\sigma_z$, best seen in face-on projections, at the same time. We
argue (\S 4.1 and Paper II) a face-on approach favoring $\sigma_z$ is
preferred for a number of reasons, but principally because little is
known empirically about the shape of disk stellar velocity ellipsoids
(SVE), while disk scale-heights are statistically well-determined from
studies of edge-on galaxies (e.g., de Grijs \& van der Kruit 1996,
Kregel et al. 2002).  Thus, $\sigma_z$ provides a direct, kinematic
estimate of $\sddisk$ and $\mldyndisk$ to break the disk-halo
degeneracy.  This, in turn allows us to unambiguously determine the
density profiles of dark matter halos.  With additional analysis of
the disk gas content and extinction, we may also constrain $\mlsdisk$ and
the IMF over a broad range of global galaxy properties local densities
and environments within each galaxy.

This kinematic approach to measuring disk mass has been attempted
before with long-slit spectroscopy, pioneered by van der Kruit \&
Freeman (1984, 1986) on several face-on and inclined systems.  This
work was the starting point and inspiration for the first survey
carried out by Bottema (1997, and references therein).
However, these observations barely reached 1.5 disk scale-lengths and
required broad radial binning. Further, the face-on samples suffered
from significant uncertainties in the inclination, and hence
total-mass estimates, while measurements of more inclined systems
required large and uncertain corrections to the observed line-of-sight
velocity dispersions ($\slos$) for the tangential
($\sigma_\theta$) and radial ($\sigma_R$) components of the SVE.

To illustrate the limitations of extant data, the spread of $B$- and
$K$-band $\mldyndisk$ in Bottema's (1993) sample is illustrated in the
bottom panels of Figure 2.  We use his formula (equation 8), which
requires a central disk surface-brightness ($\mu_0$), a central
vertical velocity dispersion ($\sigma_{z,0}$), and a disk scale-height
($z_0$). The formula and/or the $\sigma_{z,0}$ estimates assume (i)
the SVE shape is constant with radius and the same for all galaxies;
(ii) the disk scale-height and $\mldyndisk$ are constant with radius;
and (iii) the disk is purely exponential. The validity of these
assumptions are all debatable. We take velocity dispersions,
scale-lengths, and $B$-band disk surface-brightness primarily from
Bottema (1993) and earlier papers in that series, with some
supplemental surface-brightness and scale-length measurements in the
$B$ and $V$ bands compiled from the literature (van der Kruit \&
Freeman 1986; Baggett et al. 1998; McGaugh 2005). Scale-heights are
estimated from the observed scale-lengths using our calibration (Paper
II). $B-K$ colors are based on values from NASA/IPAC Extragalactic
Database (NED), corrected to total magnitudes as described in Appendix
A, corrected for Galactic extinction (but {\it not} internal
extinction), and then transformed to disk colors based on
bulge-to-disk ratios in the $B$ and $K$ bands measured as a function
of type by Graham (2001). Errors are estimated based on the quoted
observational uncertainties, differences between repeat measurements,
and, in the case of $\mldynkdisk$ and $B-K$, the dispersion in
bulge-to-disk ratios for galaxies of a given type.

Trends in Figure 2 of $B$-band $\mldyndisk$ with $B-K$ color
correlate with inclination; this is likely an internal extinction
effect. Hence the range of intrinsic $B-K$ color is small (likely less
than 1 mag). Given Bottema's small sample size and large errors,
scatter in $\mldyndisk$ -- even in the $K$ band over this color
range -- is too large to make reliable statements about disk
maximality, trends with color, or the viability of MOND. Limitations
notwithstanding, these earlier studies pointed the way forward (e.g.,
Herrmann \& Ciardullo 2009). Indeed the program we describe here is
very much a modern version of what Bahcall \& Casertano (1984)
outlined many years ago.

\section{THE DISK-MASS SURVEY}

To surmount past limitations while taking advantage of the benefits of
a direct, kinematic measurement of the mass of spiral disks, we have
built instruments, designed an analysis protocol, constructed an
observational strategy, and carried out a survey targeting nearly
face-on systems.  We refer to this as the DiskMass Survey. The range
of disk inclinations was chosen (\S 4.1) to balance uncertainties in
de-projection of both the total mass (via rotation curves) and disk
mass (via vertical velocity dispersions). This inclination choice and
the spectroscopic instrumentation built to sample kinematics in such
nearly face-on systems are the survey's hallmarks.

The DMS scope is to measure the mass surface-density in regular,
moderate-to-late type disks spanning a range in color, luminosity,
size, and surface-brightness that characterize these types of `normal'
disk systems today. The impetus to define such a sample is
several-fold. First, such a sample should include galaxies like the
Milky Way, for which we have unique (albeit limited) measurements of
the disk mass within the solar neighborhood. Second, our definition is
similar to typical TF samples that have served both as cosmological
probes and key tests of semi-analytic models of structure
formation. Third, for such a sample, bulge contributions to the
stellar kinematics are minimal outside of the inner disk scale-length.
With these considerations in mind, the range of physical properties
sampled are significant (e.g., factors of 10 to 100 in size and
luminosity). The required sample size (\S 4.2) then follows from our
intent to sample the cosmic variance in key physical parameters (e.g.,
disk-to-total mass, SVE shape, disk oblateness), as well as draw
out the principal correlations these parameters have with other
physical properties of spiral galaxies (e.g., luminosity,
surface-brightness, integrated stellar populations). The details of
the sample are described in \S 5.

The DMS experimental paradigm centers around optical bi-dimensional
spectroscopy obtained with custom-built integral-field units
(IFUs). These instruments were used to measure stellar and ionized gas
kinematics at multiple wavelengths from 500 to 900 nm, as described in
\S 4.2, covering the key spectral diagnostics of
[\ion{O}{3}]$\lambda$5007 and H$\alpha$ in emission, and \ion{Mg}{1b}
and \ion{Ca}{2} near-infrared triplet (hereafter \ion{Ca}{2}-triplet)
in stellar absorption. From these observations we are able to derive
kinematic inclinations, total mass, star-formation rates and SVEs on a
spatially resolved scale for each galaxy. The broad spectral coverage
at high dispersion was essential for providing several important
diagnostics and checks on systematics concerning our kinematic signal
due to internal extinction and variations in stellar populations
across the face of individual galaxies.

The two IFUs built for this survey were optimized, capitalizing on the
potential of two existing spectrographs, for the measurement of
two-dimensional velocity functions (e.g., centroids, dispersions, and
higher moments) of the stars and gas in spiral disks.  The salient
technical hurdle was to achieve photon-limited, medium-resolution
spectroscopy at surface-brightness levels at and below the sky
foreground continuum.  SparsePak (Bershady et al. 2004, 2005) and PPak
(Verheijen et al. 2004, Kelz et al. 2006) are large-fiber IFUs on 3.5m
telescopes (WIYN\footnote{The WIYN Observatory, a joint facility of
  the University of Wisconsin-Madison, Indiana University, Yale
  University, and the National Optical Astronomy Observatories.} and
Calar Alto, respectively) with fields-of-view of slightly over 1
arcminute. Both IFUs, illustrated in Figure 3, feed conventional
long-slit, grating-dispersed spectrographs (the Bench Spectrograph and
PMAS, respectively) configurable over a wide range of wavelengths and
spectral resolutions. The configurations relevant to the DMS are the
high-dispersion modes for \ion{Mg}{1b} (PPak and SparsePak), H$\alpha$
(SparsePak), and \ion{Ca}{2}-triplet (SparsePak) as summarized in
Table 1. Photon shot-noise is comparable to read-noise in these
configurations (Bershady et al. 2005) for dark, sky-limited
conditions. Since these IFUs are among the largest-grasp systems in
existence (Bershady et al. 2004), and grasp is conserved, their
performance serves as a cautionary note for instruments striving for
higher angular or spectral resolution, even on larger telescopes.

Near the detector-limited regime, signal-to-noise (S/N) $\propto
\sqrt{d\Omega \ \Omega}$, where $d\Omega$ and $\Omega$ are the
specific solid angle associated with each fiber and the total solid
angle of the IFU array, respectively. In all noise regimes, (S/N)
$\propto \ \sqrt{\Omega}$ for extended sources. The total solid angle of
SparsePak and PPak is 1280 and 1891 arcsec$^2$, respectively.  The
major advantage of integral-field spectroscopy (IFS) over long-slit
observations is the ability to group together fibers (i.e., coadded
spatial resolution elements) in radial bins. An example, typical of
what is used in our survey, is shown in Figure 3.  The solid angles
covered as a function of radius for each of these 5 radial bins is
plotted in Figure 4 for both IFUs. In the outer-most rings, these
instruments sample 500 to 700 arcsec$^2$, i.e., a 6th of an
arcmin$^2$.

With this formidable grasp, we are able to achieve a S/N of 12 per
pixel, or S/N $\sim 22$ per resolution element at V-band
surface-brightness of 23.2 mag arscec$^{-2}$ for spectral resolutions
(FWHM) of $\lambda/\delta\lambda \sim$ 8,000 and 11,000 (velocity
dispersions of $\sigma \sim $ 16 and 11 km s$^{-1}$) in 8 and 12
hours, respectively, for PPak and SparsePak.  These are typical
exposures for galaxies in our survey. This surface brightness is
equivalent to that of an exponential disk at 2.25 scale-lengths, given
the median central surface brightness of our sample is 21.4 mag
arcsec$^{-2}$ in the $B$ band. This is roughly 0.25 mag brighter than
a so-called Freeman disk (Freeman 1970).\footnote{For typical disk
  colors of an intermediate-type spiral, such a disk has a central
  surface-brightness, $\mu_0$, of 21.65, 21, and 20.65 mag
  arcsec$^{-2}$ in the $B$, $V$, and $R$ bands respectively.} Figure 4
also shows the delivered spectroscopic S/N in the outermost ring as a
function of surface-brightness in our typical, and our deepest,
exposures.  We obtain S/N = 10 per pixel or better at the faintest
limits of our survey, which is ample for our purposes, as we quantify
in Paper II. Figure 5 presents the coadded spectra in the outermost
radial bins for two galaxies near the extrema of the
surface-brightness range of our survey: UGC 6918 (a short exposure
with SparsePak), and UGC 1635 (a moderate exposure with PPak).  In
addition to the strong nebular line of [O~III]$\lambda$5007 and
absorption line of the \ion{Mg}{1b} triplet, the weak nebular doublet
[N~I]$\lambda\lambda$5198,5200 and a plethora of weak Fe, Ti, and Cr
absorption lines are clearly distinguishable.  We find S/N scales as
expected with exposure and coaddition.

The DMS, as a study of nearby galaxies, is distinguished by its
emphasis on (i) observations of spatially-resolved kinematics of both
gas (ionized and neutral) and stars; (ii) the direct measurement of
dynamical mass surface-density and $\mldyndisk$; and (iii) focus
on intermediate-to-late-type disk systems.  The galaxies studied were
chosen carefully to match the survey spectroscopic instruments for
efficient observation. As such, the galaxy sample is more distant
than, say, the typical galaxy in the SINGS sample (Kennicutt et
al. 2003), and more narrowly focused on normal spiral galaxies. Even
with our spectroscopic focus, photometric observations are a necessity
both to place these observations in a broader context, more completely
characterize the sample, and to properly define and interpret
$\Upsilon$ measurements in terms of stellar populations,
star-formation, and galaxy type. Accordingly, the survey as a whole
was fleshed out around our spectroscopic observations to include
ground- and space-based photometric data in the optical, near-, and
mid-infrared. Wherever possible we take advantage of existing
photometric observations in the public domain, supplementing with new
observations as necessary to increase depth or spectral
coverage. Substantial investment has been made with new observations
using the Spitzer Space Telescope (hereafter, Spitzer) as well as the
KPNO 2.1m telescope. In addition, new, 21 cm, aperture-synthesis
observations were also undertaken with the VLA, GMRT, and Westerbork
facilities to radially extend our gas-kinematic measurements and
measure \ion{H}{1} mass.

Several observational precursors were motivational in the design and
execution of the DMS. The TF study of Verheijen (1997, 2001) provided
the foundation for understanding the extent of possible variations in
the correlation of light to mass in spiral galaxies; the importance of
the near-infrared in this regard; and the subtleties of defining
rotation curve amplitudes in this analysis.  Ionized-gas kinematic
studies of face-on galaxies (Andersen 2001; Andersen et al. 2001) were
critical in understanding the power of IFS, and for developing the
techniques for deriving kinematic inclinations in the nearly face-on
regime. Finally, a pilot survey (Bershady, Verheijen \& Andersen 2002;
Verheijen, Bershady \& Andersen 2003) carried out during SparsePak
commissioning established the basic capabilities of the survey IFUs.

\section{SURVEY STRATEGY}

\subsection{The Nearly Face-On Approach}

Because the two primary observables contributing to the estimation of
$\sddisk$, namely $\sigma_z$ and $z_0$, have orthogonal
projections to the observer's line-of-sight, the choice of galaxy
inclination is paramount in optimizing the error budget. Intermediate
inclinations proffer access to both $\sigma_z$ and $z_0$, but in
practice the task is difficult: $\sigma_z$ is quickly overwhelmed by
the other SVE components if, as expected, $\sigma_z$ is the smallest
component; non-axisymmetric photometric features in disks associated
with spiral structure (e.g., dust, star-formation) preclude a simple
estimation of disk thickness at all but nearly edge-on
orientations. Consequently, compelling survey strategies either focus
on edge-on galaxies, where the vertical scale-height can be directly
observed, or on nearly face-on galaxies where $\slos$ is
dominated by the $\sigma_z$ component. The projection of $\sigma_z$
into the line-of-sight along a galaxy's kinematic minor axis is shown
in Figure 6 as a function of inclination and SVE shape. In the
edge-on case $\slos$ contains nothing of $\sigma_z$, and hence
one must rely on knowledge of the SVE to extract this vertical
component. Conversely, in the face-on case one must rely on knowledge
of the typical vertical scale-height of the kinematically-probed
stars, or a re-scaling of the radial scale-length based on an adopted
disk oblateness.

The goal of our survey strategy is to be limited by the uncertainties
in the disk vertical scale-height of galaxies, specifically the
dispersion in the relation between $z_0$ and observables that do not
require edge-on orientation (Paper II). In other words, we opt for the
``face-on disk'' approach. Our rational is several-fold. Foremost is
the fact that $\sddisk$ depends quadratically on $\sigma_z$, but
only linearly on $z_0$. Furthermore, we know far less about the shape
of the disk SVE than about disk photometric oblateness.  As we
quantify in Paper II, our ability to estimate disk oblateness in
face-on galaxies is 2-4 times better than our ability to estimate the
SVE in edge-on systems.  This translates into a 4 to 16 times increase
in precision for measuring disk-mass in the face-on regime.

On top of these primary considerations, given the complications of
projection and extinction in edge-on galaxies -- for both kinematics
and photometrics -- clearly the face-on approach is desirable in terms
of minimizing systematics. Since surface-brightness falls off
exponentially in disks, the past observational challenge has been to
gather enough light in the outer regions of disks to make a
sufficiently precise kinematic measurement. Edge-on galaxies
conveniently project the disk light into a geometry suitable for a
long-slit spectrograph. However, with our development of appropriately
sized IFUs, we can fully sample face-on disks, and optimally average
in azimuth and radius as our detailed analysis requires. The coupling
of a nearly face-on approach with IFS means that systematics can be
minimized without enlarging our random errors -- again, the hallmark
of our approach. Inclinations close to 30 degrees are optimal for
determining the disk-mass surface-density simultaneously with the
total galaxy mass (Paper II), and conveniently are also optimal for
measuring the SVE (Westfall 2009). As shown in Figure 6, this is where
we expect the radial and vertical components of the SVE to equally
project into the line-of-sight.

\subsection{Scope and Protocol}

Our aim is to sample the disk dynamical and stellar mass-to-light
ratios ($\mldyndisk$ and $\mlsdisk$) of intermediate-to-late type
spiral galaxies spanning a range in luminosity, surface-brightness,
and color, but with only a modest range in other morphological
attributes. For this sample, we also aim to know the gas content and
star-formation rate on a spatially resolved scale commensurate with
the measurements of disk $\mldyndisk$. With 3 bins in each parameter
(luminosity, surface-brightness and color), a sample of $\sim$40
galaxies is needed to have a sufficient number of galaxies in each
bin. This is not simply to diminish our random errors, but also to
limit our systematic errors, as discussed in Paper II, and to
understand the range of astrophysical variance in $\mldyndisk$ and
$\mls$.  These aims set the basic scope of our survey.

\subsubsection{Down-Selecting}

The spectroscopic requirements of the survey are demanding, with the
stellar absorption-line observations by far the most taxing.  To make
the survey data acquisition tractable, we designed a two-phase
protocol following a down-selecting scheme. The protocol focuses
resources on a modest sub-sample of galaxies, yet provides a large
enough parent sample to place our study into the context of
disk-galaxy properties and scaling relations. The selection criteria
and sample details for each of these phases are given in \S 5.

The initial survey phase, Phase-A, is based on a purely photometric
selection of a large number of targets that suit our observational and
scientific criteria. After detailed inspection, roughly 14\% of this
parent sample was deemed suitable for spectroscopic investigation of
their ionized-gas kinematics. H$\alpha$ velocity fields for 63\% of
these sources were successfully obtained (about 9\% of the parent
sample defined in \S 5.1). Ground-based optical and near-infrared
observations targeted this subset. Observations in this phase had the
immediate purpose of enabling us to estimate inclination, total mass,
and mass-luminosity scaling properties; the additional purpose of
establishing their kinematic regularity for Phase-B selection; and the
final purpose of estimating star-formation rates and diagnostics
parameters of the inter-stellar medium(ISM) such as line-widths and
ratios of the ionized gas.

In the second phase, Phase-B, we further down-selected $\sim$32\% of
the spectroscopically-observed Phase-A sample for intensive further
study, including spectroscopic observations of their stellar
kinematics, aperture-synthesis observations of their \ion{H}{1}
distribution, and Spitzer observations of their mid-IR flux
distributions. This core sample amounts to $\sim$20\% of the initial
Phase-A sample, and only 3\% of the parent sample.

\subsection{Spectroscopic Coverage}

Spectral regions were chosen to characterize neutral- and ionized-gas
content and kinematics via emission, and stellar kinematics via
absorption-line observations.

\subsubsection{Gas kinematics}

For the Phase-A sample, the H$\alpha$ region was selected because of
the strength of the Balmer line; the ability to sample other, strong
nebular lines ([\ion{N}{2}]$\lambda\lambda$6548,6583 and \newline
[SII]$\lambda\lambda$6716,6730) even in the small spectral range
sampled at medium resolution; the utility of H$\alpha$ as a
star-formation-rate (SFR) indicator; and finally because of the
relative efficiency of obtaining velocity fields using SparsePak on
the WIYN Bench Spectrograph. Spectroscopic observations are described
in \S 6.1.

For the Phase B sample, spatially-resolved \ion{H}{1} observations
were a critical augmentation because the observable \ion{H}{1} gas
disks typically extend well beyond the field-of-view of the SparsePak
IFU with which the H$\alpha$ velocity fields have been mapped and from
which the inner rotation curves have been determined.  From the more
extended \ion{H}{1} velocity fields, the rotation curve can be
measured well into the outskirts of the galaxies where the dynamics
are dominated by the dark matter halo.  This allows us to properly
measure the radial density profiles of the dark matter halos from an
unambiguous rotation curve decomposition.  Extended \ion{H}{1}
velocity fields and rotation curves are also required for detailed
Tully-Fisher studies.  Aperture-synthesis observations at 21 cm with
the VLA, WSRT and GMRT are described in \S 6.1.2.

Our 21 cm observations also serve to constrain the gas-mass budget.
By measuring the \ion{H}{1} column-density distribution in the disk,
we determine the total (H+He) gas surface-density. This is essential
for properly calibrating the {\it stellar} mass-to-light ratio of the
disk, $\mlsdisk$, especially in the outer regions. The neutral
gas-density is also of interest in relation to SFR indicators (e.g.,
H$\alpha$, 24 $\mu$m flux, etc.) to study star-formation thresholds
and gas-consumption time-scales, and hence better link observed
stellar populations with their star-formation histories and
potentials.

\subsubsection{Stellar kinematics}

Several considerations led to the definition of the stellar-kinematics
observations. Primary was to obtain a measure of the kinematics of the
old stars that are dynamically relaxed, and therefore well-sample the
disk potential, and are representative of the stellar population
dominating the vertical light distribution in red and
near-infrared light of edge-on samples.
Secondary was to sample, if at all possible, both the stellar and ionized-gas
kinematics simultaneously, in one spectroscopic-instrumental setting.
Overall, the instrumental configurations were required to deliver
spectral resolutions above $\lambda/\delta\lambda = 8500$ ($\sigma<15$
km s$^{-1}$) and have reasonable efficiency.

Our primary consideration for the stellar-kinematic measures is
critical because it is well known that the scale-height and velocity
dispersion of stars in the Milky Way are correlated with their
spectral type, and presumably their mean age. The physical picture is
that stars, born in the disk mid-plane, slowly diffuse and dynamically
relax with time. This picture should apply generally to other disk
systems, and hence our measurement scheme must be sensitive to, or
avoid systematics in disk-mass measurements due to this physical
effect. This effect would include, for example, the impact of an older
thick-disk component on both the effective (light-weighted) vertical
scale-height and $\slos$.

Historically, the spectral regions of interest for stellar-kinematic
work has stemmed from a focus on early-type galaxy observations. Since
such galaxies do not suffer from hot-star contamination, the strong
H\&K lines near 400 nm, in combination with the ubiquitous and
intrinsically narrow Fe lines in the blue have been a favored target
well matched to optical spectrograph performance and low sky
foregrounds. In galaxies with composite stellar populations, as in
our survey, the picture is necessarily more complicated. The 400-nm
region is too blue to have sufficient sensitivity to the old stellar
population in our survey galaxies.

To illustrate the issues of sampling the old stellar disk population,
Figure 7 shows a simple, 3-star spectral decomposition of
intermediate-type spiral galaxy with colors typical of our sample.
The decomposition is motivated by the earlier work of Aaronson (1978)
and Bershady (1995), and in fact uses the specific stars Bershady
found most-frequently provided best fits to galaxy colors of this
intermediate (spectral) type. While the fit is not unique, it provides
an excellent match to the observed optical to near-infrared broad-band
colors, and is representative of the types of stars which dominate the
integrated light at different wavelengths. The three stars (B9~V,
K0~III, and M5~III) represent young, old, and intermediate-age stellar
populations, respectively. Ideally it is the K-giant population we
would like to sample. Below 500 nm, hot young stars dominate; above
500 nm K-giants dominate, but at near-infrared wavelengths there is an
increasing contribution from intermediate-age M-giants.

Two facts provide nuance to this picture. Hot stars, outside of the
Hydrogen and Helium lines, have weak spectral features, and so their
contribution to the kinematic signal is effectively to diminish the
amplitude of the old-star signal, but not to modulate significantly
the velocity structure. However, late-A and F-stars, also
characteristic of intermediate-age populations do contribute some
signal in metal absorption lines (see \S 6.2.2); these stars, if in
sufficient abundance, may modulate the kinematic signal in the
visible. Consequently, the blue-visible region may also have
significant contribution from intermediate-age stars. These
considerations drive us to work in the 500-1000 nm region, i.e.,
farther in the red, but not too far into the near-infrared to become
dominated by potential intermediate-age TP-AGB stars or red
super-giants in vigorously star-forming regions.

In the 500-1000 nm range, popular regions for stellar kinematic study
have included the \ion{Mg}{1b}-triplet region near 513 nm and the
\ion{Ca}{2}-triplet near 860 nm, with the latter enabled by
red-sensitive CCDs. Studies using the \ion{Ca}{2}-triplet region have
the advantage that a single gravity-insensitive ion dominates the
cross-correlation signal. Therefore one would expect it simpler to
match the temperature and metallicity between stellar templates and
the target galaxy, thereby minimizing systematic error in the derived
kinematics. In contrast, Barth, Ho \& Sargent (2002) point out the
potential difficulties in the \ion{Mg}{1b}-triplet region due to a
combination of ions whose strength vary with both temperature,
metalicity {\it and} relative abundance. We find (Paper II) that the
dependence of the derived $\slos$ on template temperature is
comparable in amplitude, although different in trend in these two
regons. One of the clear disadvantages of the \ion{Ca}{2}-triplet
region for measurement of velocity dispersions in dynamically cold
systems is the great strength of the lines, which therefore are
intrinsically broad. Bershady et al. (2005) measured intrinsic widths
of the \ion{Ca}{2}-triplet between 25-35 km s$^{-1}$ ($\sigma$) in
cool stars in the context of cross-correlation measurements with
SparsePak, directly relevant to the study at hand. They found the
\ion{Mg}{1b} triplet (513 nm) has similar intrinsic widths to the
\ion{Ca}{2}-triplet, but is surrounded by weak iron lines with
intrinsic widths of only 7-8 km s$^{-1}$. For disk systems, where
dynamical widths are expected to be of order 10's of km s$^{-1}$,
narrow (and therefore weak) lines are essential. This points to a
final desideratum, namely we would like a region where there are some
strong lines that can be used for velocity centroiding (e.g., spectral
stacking, \S 5.2), but many weak lines so there is significant signal
from intrinsically narrow features (in the absence of velocity
broadening from the galactic potential).

The \ion{Mg}{1b} region is particularly good in this regard, and
offers three other advantages relative to the \ion{Ca}{2}-triplet and
other regions: (i) the presence of
[\ion{O}{3}]$\lambda\lambda$4959,5007 doublet within a spectral window
small enough to contain key absorption-line features measurable at
high-dispersion; (ii) low sky-foregrounds, and in particular the
absence of strong sky line-emission; and (iii) relatively good
instrument performance. (A comparison of the WIYN Bench Spectrograph
performance in \ion{Mg}{1b} and \ion{Ca}{2}-triplet regions is given
in Bershady et al. 2005; in the case of PMAS, no grating was available
with adequate resolution in the \ion{Ca}{2}-triplet region). While an
intermediate-wavelength region, such as 600 nm might be ideal from a
stellar-population perspective, and weak lines do exist there, no
strong absorption or nebular line-emission is present. The H$\alpha$
region, just slightly to the red, is devoid of sufficient
absorption-line signal.  Our primary region for measuring
stellar-absorption was then adopted to be the \ion{Mg}{1b} region.
However, multiple wavelengths permit independent determinations of our
primary observable, the SVE, thereby providing a critical cross-check
on systematics due to cross-correlation template-mismatch or
astrophysical variance in scale-heights and velocity dispersion with
stellar type. Accordingly, we observed with SparsePak a subset of
galaxies in the \ion{Ca}{2}-triplet region as a cross-check on our
\ion{Mg}{1b} observations.

\subsubsection{Spectral Library}

One critical issue for estimating $\slos$ accurately is
determining suitable templates for deriving the broadening function.
The so-called ``template mismatch'' problem arises when the derived
broadening function is systematically in error due to the unsuitable
match between stellar template and galaxy spectrum. Template-mismatch
is thought to be relatively minor for early-type galaxies observed in
the blue-visible (400-500 nm) portion of the spectrum and for any
galaxy observed in the \ion{Ca}{2} region. For the former, the approximation
of early-type galaxies to simple stellar populations means a single
stellar template best representing the tip of the red giant branch
(e.g., G8 III to K1 III) is likely a good first-approximation. For the
latter, the intrinsic width of the \ion{Ca}{2} lines which dominate the
kinematic signal in this spectral region are weakly dependent on
spectral type.

Because the DMS focuses on intermediate-type spirals with on-going
star-formation, these simplifying assumptions are not valid. Not only
are the DMS spirals clearly composite (young+old) stellar populations,
but given the uncertainties in the TP-AGB life-time, contributions from
very cool stars may also be significant, particularly in the \ion{Ca}{2}
region. The impact of this complexity is illustrated in Figure 7.  In
light of the potential for independent thin disk / thick disk
kinematics that correlate with stellar population age and metallicity,
an application of the simultaneous multi-component kinematic study of
de Bryne et al. (2004) to spiral disks may be necessary.

For these reasons, a large template library that spans temperature and
metallicity is important for our program. When we began the survey,
there were no population synthesis models available with spectral
resolution comparable to our data. While this has begun to change
(e.g., Le Borgne et al. 2004), it was observationally inexpensive to
gather our own library (\S 6.2.2).  These observations allow us to
synthesize stellar populations while at the same time minimizing
systematic errors due to instrumental signatures present or absent in
model spectra. This is an important issue that we return to in Paper
II.

\subsection{Photometric Coverage}

\subsubsection{Optical--near-infrared wavelengths}

Existing photometric data for all galaxies in the DMS consist of
optical DSS\footnote{The Digitized Sky Survey, based on POSS-II
  plates, was produced at the Space Telescope Science Institute under
  U.S. Government grant NAG W-2166.} blue and red images, as well as
2MASS (Skrutskie et al. 2006) near-infrared images and photometry,
along with on-line compilations, e.g., NED, from which we extract
$B$-band photometry from RC3 (de Vaucouleurs et al. 1991) and
integrated \ion{H}{1} fluxes from a variety of sources.  These data
are sufficient for a preliminary sample selection (\S 5). Over the
course of the survey, the Sloan Digital Sky Survey (SDSS; York et
al. 2000) has released data covering about 40\% of the sample. While
SDSS images are deep and flat enough for our analysis purposes, the
incomplete coverage is problematic.

Complete, multi-band imaging data is required to deliver measurements
of luminosity, color, surface-brightness, and scale length -- vital
components for measuring $\Upsilon$, carrying out mass decompositions,
constraining stellar population models, and studying the Tully-Fisher
relation. The wavelength coverage must overlap explicitly the
wavelength regions where we have high-resolution spectroscopic data
(for the pragmatic purposes of registering our IFS data), but extend
to the near-UV and near-IR to be sensitive to, and constrain both
young and old stellar populations. Therefore, our spectroscopic data
were augmented with new, ground--based optical ($UBVRI$) and
near-infrared ($JHK$) photometry, as described in \S 6.3.1. These data
extend and deepen the existing archival optical and infrared imaging
from SDSS and 2MASS.  Imaging campaigns focused on the Phase-B subset,
but cover as much of the full Phase-A sample as possible.

\subsubsection{Mid-infrared wavelengths}

Spitzer near and mid-infrared (4.5, 8, 24, and 70 $\mu$m) images were
taken of the majority of the Phase-B sample. Central themes of Spitzer
observing programs include the uncovering of obscured star formation,
reliable assessment of bolometric luminosity, and the determination of
how star formation depends on properties of the interstellar medium.
Infrared surface photometry between 5-100 $\mu$m of nearby galaxies is
sensitive to cold and warm dust emission, and allows for this emission
to be characterized and differentiated from stellar emission.  Such
photometry thereby permits a determination of the extinction and
bolometric star formation rate. By making such measurements for the
DMS, particularly for the sample with \ion{H}{1} imaging and H$\alpha$
spectroscopy, we tie extinction, dust properties and bolometric star
formation rates directly to both dynamical and gas-mass
surface-density. While this connection is not essential to
mass-decomposition per se, the fundamental connection between how well
light traces mass, and how star formation traces mass surface-density
are intimately linked. Our motivation for Spitzer observations are to
make these connections, extend the calibration of stellar
mass-to-light ratios to as long a wavelength as possible, and relate
star formation to the physical environment within galaxies --
including the local gravitational potential.

\section{SURVEY SAMPLE}

\subsection{Source Selection}

Since the initiation of this survey pre-dated the release of
significant portions of SDSS, we relied on the UGC (Nilson, 1973) and
DSS images for source selection. The UGC is complete for sources
with blue Palomar Sky Survey m$_{pg} < 14.5$ (roughly $B = 14.7$)
and/or diameters $>$ 1 arcminute. The Phase-A source list was selected
to contain morphologically regular intermediate-to-late-type disk
galaxies, with low apparent photometric inclinations, modest to little
foreground extinction, and the right apparent size for our IFUs.
Regular morphology was desired to ensure regular gas and stellar
velocity fields to aid in measurement of disk inclinations, and
simplify interpretations of stellar kinematics.  In short, the
selection resulted in a sample similar to that of Andersen et
al. (2001; see also Andersen 2001), which was a fore-runner to this
survey. However, the apparent sizes of the galaxies are slightly
larger on average to fit the SparsePak field-of-view, and we were less
restrictive here on barred galaxies; systems with small and/or weak
bars are part of our survey.

To cull a sample for spectroscopic follow-up, we first selected a
subset of the UGC as a {\it parent} sample based on the specific
criteria of UGC-tabulated values of blue diameter between $1^\prime <
D_{25} < 3.5^\prime$ and blue-band minor-to-major axis ratios
$b/a\geq 0.75$ (nominally inclinations under 41.4$\circ$), plus the
RC3-tabulated values of Galactic extinction A$_B^{\rm g} < 0.25$ mag and
numerical Hubble type $T \geq 0$. We avoided heeding existing, qualitative
morphological classification (although interesting to review a
posteriori) except in this one cut to remove the earliest-type
galaxies. This yielded 1661 sources.

For these sources we carried out our own, isophotal, ellipse-fitting
surface-photometry on red DSS images, calibrated with the existing
surface-photometry of Swaters \& Balcells (2002). Given the limited
quality, depth, and dynamic range of the DSS data we did not pursue a
quantitative classification (e.g., Bershady et al. 2000), but instead
used these measurements to estimate disk central
surface-brightness, scale-length, and ellipticity. An automated
fitting procedure was carried out between $R$-band surface-brightness
levels of 21.5 and 24.5 mag arcsec$^{-2}$. The bright limits was
chosen to exclude portions of the profile where photographic
saturation was an issue. The faint limit was estimated on the basis of
the profiles to correspond to the depth below which uncertainties in
sky subtraction were significant.  The images were prepared by
removing stars automatically via a spatial filter (un-sharp
masking). The center of each galaxy was excluded from the filtering.
The information from this fitting procedure was used to further
down-select from the parent sample based on two quasi-independent
schemes.

One scheme involved four independent visual inspections of the light
profiles and images to select objects of appropriate size and
inclination, while eliminating objects with gross asymmetries, large
and strong bars, large bulges, very low surface brightness disks,
nearby neighbors, and bright stars. A sample of 176 sources were
agreed upon by all 4 reviewers as satisfactory targets.  We refer to
this as the 4/4 sample.

A second scheme aimed at a more quantitative version of this same
exercise, consisting of the following three steps: (i) Select galaxies
with $25^{\prime\prime} < r_{23.5} < 45^{\prime\prime}$, where
$r_{23.5}$ is the radius at which $\mu=23.5$ mag arcsec$^{-2}$ in the
$R$ band; (ii) eliminate strongly-barred, interacting, or peculiar
galaxies, or galaxies with bright stars in their optical diameters --
again, based on visual inspection; and (iii) select galaxies with
$10^{\prime\prime}<h_R<20^{\prime\prime}$.  The combination of the
first and third cuts selects galaxies with roughly 3$h_R$ within the
SparsePak field of view for high-surface-brightness galaxies and
2$h_R$ for lower surface-brightness galaxies. In total this yielded
128 sources. We refer to this as the DSS sample.  Of these, 98\% were
also selected by two or more reviewers based on visual selection
alone, and 66\% were selected by all four reviewers.

The nominal Phase-A sample, deemed suitable for H$\alpha$
spectroscopic follow-up, consisted of the union of the 4/4 and DSS
samples (219 sources). To this we added 12 objects with similar
properties with existing H$\alpha$ IFS obtained as part of our
37-galaxy pilot survey (25 targets were already included in the 4/4 or
DSS samples). The pilot-survey sources were selected from de Jong
(1994), the NFGS (Jansen et al. 2000), Andersen et al. (2006), or an
extension of the Andersen et al. sample to larger sizes -- in all
cases similar criteria as applied to the 4/4 or DSS samples.  The
final Phase-A sample, then contains 231 galaxies, all from the UGC.

Of the full Phase-A samples, H$\alpha$ IFS was obtained
for a total of 146 galaxies, based on the available observing runs,
target visibility, and any additional assessment of target suitability
by the observer, based on the photometric selection desiderata. A
subset of 124 galaxies are nominally deemed to have high-quality
data. This H$\alpha$ subset forms the portion of the Phase-A sample
from which the Phase-B sample is selected. Henceforth we refer to this
subset as the H$\alpha$ sample.

While we relied on apparent morphology for the initial Phase-A sample
selection, in contrast the Phase-B sample selection used the
additional information from H$\alpha$ velocity fields. Our aim was to
select $\sim$40 spiral galaxies for stellar absorption-line
spectroscopy follow-up with regular velocity fields and
well-determined, but low inclinations consistent with the discussion
on errors in total and disk mass in Paper II. Based on our pilot
survey we expected 40\% of our H$\alpha$ sample would meet this
criterion. In practice, we found that only 25-30\% out of our full
H$\alpha$ sample were suitable, and this drove us to increase our
H$\alpha$ survey of Phase-A targets from 100 to 146 sources.

To accommodate sources spanning as large a range in color, luminosity,
and surface-brightness we relaxed our Phase-B selection. In one
notable case (UGC 4256) we specifically included a source with strong
spiral perturbations to ascertain the impact of non-axisymmetric gas
motions on our disk-mass estimation. This source has strong
star-formation rate, blue colors, and high surface-brightness. This
set of 40 Phase-B galaxies was observed with Spitzer. The final
Phase-B sample was augmented with an additional 6 galaxies in the
H$\alpha$ sample with stellar absorption-line IFS obtained over the
course of the survey. Of these 46 galaxies, 44 have high-quality
H$\alpha$ IFS and 42 have high-quality stellar IFS. UGC 6903 is the
only galaxy observed with Spitzer for which our stellar IFS data have
not met our S/N requirements.

A summary definition of the survey sample, inclusive of our full
Phase-A source list, is in Table 2.\footnote{The following additional
  objects were observed as part of our pilot program as calibration
  sources primarily for our stellar velocity-dispersion measurements:
  UGC 6869, an inclined, high surface-brightness compact spiral in
  Ursa Major; UGC 11012, an inclined spiral observed by Bottema
  (1989); and three ellipticals UGC 11356, UGC 5902, and UGC 9961,
  with published stellar velocity dispersions.}  Coordinates (col. 2
and 3) specify the nominal centers of our IFS observations and Spitzer
pointings, where relevant.  Morphological types (col. 4) rationalize
the independent UGC and RC3 designations\footnote{UGC and RC3
  essentially always agree on the `family' designation `SB' (barred);
  we adopt the `SAB' designation from RC3 and drop `S', neither of
  which are in the UGC classification.  We also drop the `variety'
  from RC3 (ringed/non-ringed) although this will be of interest later
  in our dynamical studies.  RC3 Hubble-types for spirals are often a
  half-type later than for the UGC; we average between the two where
  relevant.  We also adopt the RC3 Sdm/Im nomenclature. Unusual
  properties are flagged, as noted in the table.}.  Heliocentric
velocities ($V_{\rm hel}$) and distances (D) are adopted from NED
(col. 5 and 6); distances use flow-corrections for Virgo, the Great
Attractor, and Shapley Supercluster infall. The $B$-band Galactic
extinction (col. 7) is based on IRAS measurements (Schlegel et
al. 1998), as listed by NED.  Apparent $B$-band photometry from RC3
and $K$-band photometry from 2MASS are listed in cols. 8 and 9.  These
are total magnitudes corrected for systematics which become
appreciable beyond $B = 13$ and $K = 10$ mag (see Appendix
A). Absolute $K$-band magnitudes and rest-frame $B-K$ colors (cols. 10
and 11) are corrected for Galactic extinction and distance-modulus
only. Our measurements from red DSS images of disk central
surface-brightness ($\mu_{0}$) calibrated to the $R$ band, isophotal
radius ($r_{23.5}$), disk scale-length ($h_R$), and apparent
ellipticity are in cols. 12-14. Column 15 indicates how a source was
selected to be in the Phase-A sample. The criteria most important for
selection in the Phase-A sample are A$_B^{\rm g}$, $h_R$ and
$r_{23.5}$, in columns 7, 13, 14, respectively.  Columns 16-19 report
relevant observations for Phase-A and Phase-B portions of the survey,
including H$\alpha$, \ion{Mg}{1b}-region or \ion{Ca}{2}-region
spectroscopy, Spitzer imaging, and \ion{H}{1} imaging. Sources with
stellar spectroscopy, \ion{H}{1}, or Spitzer imaging are part of the
Phase-B sample.  The distribution of Hubble types, barred,
weakly-barred, and un-barred for both Phase-A and Phase-B samples is
shown in Figure 8.  The Phase-B sample observed with Spitzer,
representative of our full sample, is shown in Figure 9.

\subsection{Sample Characterization}

We make a provisional photometric characterization of our Phase-A and
Phase-B samples in Figures 10-14 based on the tabulated redshifts, $B$
and $K$-band photometry corrected from the literature, and our central
disk surface-brightness measurements. A numerical summary of
properties is given in Table 3.  Figure 10 shows the Phase-A and
Phase-B samples have recession velocities of 200-14,000 km/s and $B <
14.7$ (the UGC completeness limit). This corresponds roughly
to $K < 12$. The one source (UGC 3965) fainter than the UGC $B$-band
limit is a broad-lined AGN, and presumably faint in the RC3 photometry
due to variability. Distances range from 1.3 to 200 Mpc for the
Phase-A sample and 18 to 178 Mpc for the Phase-B sample, although 90\%
of all these samples are contained between 19 and 135 Mpc, with median
distances of 62 and 65 Mpc for the Phase-A and Phase-B samples,
respectively.

While our Phase-A and Phase-B samples all lie within the nominal
completeness limit of the UGC, the samples are not strictly
magnitude-limited.  This is illustrated in Figure 11, which shows the
$B$-band counts for the parent, Phase-A, and Phase-B samples.  While
the parent sample shows a close-to-Euclidean slope of 0.6 dex for
$B<14$, the other samples are substantially sub-Euclidean. In
otherwords, our selection of Phase-A and Phase-B samples has
preferentially excluded the fainter sources from the parent sample.
Nonetheless, like magnitude-limited samples, both the Phase-A and
Phase-B luminosity distributions both peak in $B$ and $K$ bands near
$M^*$, truncating quickly $\sim2$ magnitudes brighter than this peak,
but with extended tails to lower luminosities.\footnote{$M^*$ is the
  knee in the luminosity function with values in the $B$ and $K$ bands
  of $M^*_B = -20.6$ and $M^*_K = -24.6$ (see, e.g., Bershady et
  al. 1998 for the $K$ band). The associated luminosity is $L^*$.  The
  superscript here does {\it not} denote `stellar.'}  The median
luminosity values also vary little between sub-samples, in the range
of -20.5$\pm0.1$ and -23.8$\pm0.2$ for$M_B$ and $M_K$
respectively. The full luminosity range of the Phase-A sample spans a
factors of $10^4$ in $L_B$ and $L_K$, a large fraction of which is due
to one extreme low-luminosity source (UGC 7414) at low heliocenteric
velocity. This source has relatively large flow-corrections, and hence
its distance and luminosities are uncertain.  A more robust
characterization is based on the luminosity range enclosing 90\% of
the Phase-A sample; this spans factors of $>$40 in $L_B$ and $>$150 in
$L_K$. The full range of luminosties for the Phase-B sample, which
does not include UGC 7414, span factors of $\sim$55 in both $L_B$ and
$L_K$. In short, the sample spans about a factor of 100 in $L_K$,
which should correspond to a comparable range in stellar mass.


The optical--near-infrared color-luminosity distribution in Figure 12
shows the DMS spans a wide range of disk types. Both the Phase-A and
Phase-B samples' colors range from $2.0 < B-K < 4.25$, or about a factor
of 8 in $L_B/L_K$. The subset with \ion{Ca}{2}-region IFS
spans only a factor of 4 in $L_B/L_K$, but is well dispersed within
the color-luminosity distribution. The range of rest-frame colors
spans galaxy spectral types (Bershady 1995) from bk through gm, i.e.,
galaxies dominated by the light of B+K stars through G+M stars. The
regressions shown in Figure 12 have been converted to our adopted
cosmology and photometric band-passes. For example, consistent with
the analysis presented in \S 4.3.2 (Figure 7), UGC 6918 is a bm-type
galaxy, close to the boundary for bk-type galaxies, i.e.,
well-represented by a 3-star admixture of B+K+M stars.  The wide range
of spectral types indicates a wide range of stellar populations are
present in the sample, well-suited for probing variations in
$\mls$ and their putative correlations with color.

While the DMS sample is not complete in any quantitative sense, it
samples well the population of luminous spiral galaxies in the nearby
field. The sample color-luminosity distribution matches the
trends seen in deeper fields samples, e.g., the $B<20.5$ sample from
Bershady et al. (1994), but is absent the low-luminosity (dwarf)
systems ($M_K>-20.5$) at blue color ($B-K<3$).  Consequently, the
distribution follows the ridge-line for spirals seen in brighter
surveys (labeled ``Spirals'' in Figure 12) corresponding to $M^*$ that
decreases with later galaxy types.

There are the 17 galaxies within 0.2 mag (or redder) of the
red-sequence (labeled here gm or ``Spheroids''). By definition of the
sample, this is not due to inclination-effects. Of these, 12 galaxies
are morphologically classified as Sab or earlier, i.e., bulge, or
spheroid-dominated systems. One lying well above the red-sequence is
the broad-lined AGN noted above. Of the remaining 5 galaxies, 3 are
classified only as ``S'', but visual inspection reveals systems with
either a large bulge (UGC 7205), of very early type (UGC 9610; a
barred S0), and possibly a dusty star-burst (UGC 12418). The other two
galaxies are classified as Sb or Sbc systems, lie slightly below the
red sequence, and hence are consistent with the tail of the
intermediate-spiral distribution.

Finally, the disk central surface-brightness distribution is shown in
Figure 13 versus $B-K$ color and $M_K$. The well-known but weak trends
to lower surface-brightness at bluer color and lower luminosity are
seen. Size and luminosity are tightly correlated in the sample (about
a factor of 3 range in size at given luminosity; Figure 14), while
size and surface-brightness are not. The full Phase-A and Phase-B
samples span a factor of 35 in surface-brightnesses about the Freeman
value. The surface-brightness range enclosing 90\% of these samples
span 3.5 times higher and two times lower than the Freeman value.  In
general, the subset of Phase-B galaxies well-samples the full Phase-A
distribution, except at luminosities $M_K>-21.5$. The survey as a
whole selects against low surface-brightness disks. This is a result
of our size limit coupled with the depth limitations of the UGC. The
resulting samples are well matched to the observational capabilities
of current instrumentation, and still probe a wide range in disk
surface-brightness. In comparison (cf. Table 3), Bottema's (1997)
sample spans a similar range in surface-brightess ($\mu_0 =
20.25^{+1.2}_{-0.8}$ in the $R$ band), color ($B-K =
3.5^{+0.8}_{-0.5}$), and luminisity ($M_K = -23.6^{+1.9}_{-1.5}$).

Given the complex structure and large dynamic range of spiral-galaxy
light-profiles, there were significant instances where the automated
surface-photometry fitting described in \S 5.1 failed to provide an
adequate description of the disk radial and luminosity
scales. Typically this was because a light profile showed evidence for
inner breaks (type I and II luminosity profiles; Freeman 1970) and/or
outer breaks (Pohlen \& Trujillo 2006). In these cases we redid the
fitting after identifying the radial region outside of an inner break
corresponding to the extent of the SparsePak footprint. The disk
central surface-brightness and scale-length values in the figures and
Table 2 reflect values from the updated fits. Even these updated fits
are only an exponential characterization of the light profile within
the region where we have measured gas and stellar kinematics. In the
table we flag the three cases where the updated values place the
source outside of the DSS-sample selection definition.

\section{OBSERVATIONS}

\subsection{Total and Gas Masses}

The kinematic data used to dynamically determine the total mass
distribution come from two-dimensional velocity-fields measured in
H$\alpha$ and \ion{H}{1}. We summarize the basic observations, and
indicate how the data are used.

\subsubsection{H$\alpha$ Integral-Field Spectroscopy}

H$\alpha$ kinematic data were taken with the SparsePak IFU in the
echelle configuration specified in Table 1. We observed 137 galaxies
over 13 runs from January 2002 to April 2005 totaling 41.5 nights, plus
portions of two additional runs totaling 8 nights during SparsePak
commissioning in May-June 2001. As noted above, 14 galaxies in the
survey had prior observations using DensePak, a similar IFU, and
essentially the identical spectrograph configuration. Eight of these
were not re-observed with SparsePak.

Relevant science data include a 3-position dither-pattern for
H$\alpha$ to create a filled-in map, although in some cases this was
not achieved.  With a completed, 3-position pointing, the fill-factor
is $\sim$65\%.  Galaxies without 3 pointings or observed in poor
conditions are flagged as ``low quality'' in Table 2, regardless of
the detected signal level. Of the 20 galaxies (13\%) with low-quality
H$\alpha$ data, only four are in the Phase-B sample, and only three
are in the Spitzer sub-set.  Regardless, the H$\alpha$ data available
are sufficient for velocity-field modeling of all Phase-B sources.
The typical extent of the velocity field reaches to between 2 and 4
disk scale-lengths.

The basic analysis of the H$\alpha$ spectra consists of preliminary
processing, characterization of all the nebular lines in the echelle
order (H$\alpha$, [\ion{N}{2}]$\lambda\lambda$6548,6583, and often the
sulfur doublet [SII]$\lambda\lambda$6717,6730), and estimating the
spectral continuum. Preliminary processing (extraction, rectification,
and calibration) was carried out with standard routines found in
IRAF. Minor parameter modifications particular to the SparsePak data
format, as well as alternative sky-subtraction methods are detailed
elsewhere.  Line-fitting is done with a custom-built code which fits
single and double Gaussians to each profile, deriving line strengths,
centroids, and widths, as well as accurate errors (Andersen et
al. 2006, Andersen et al. 2008). Lines with strong skew, bimodality,
or unusually large widths are identified.

These basic data products allow us to measure inclination, rotation
speed, and kinematic regularity using velocity fields generated from
line centroid maps; to spatially register the bi-dimensional
spectroscopic data both in a relative and absolute sense (Paper II)
using continuum and velocity maps; to estimate star-formation rates
from H$\alpha$ equivalent widths calibrated with broad-band
photometry; and to crudely estimate metallicity using
[\ion{N}{2}]/H$\alpha$ line ratios. These results will be part of the
DMS paper series. A full description of the observations, reduction
and basic analysis of the emission-line profiles are presented in
Swaters et al. (2010, in preparation) and Andersen et al. (2010, in
preparation).

\subsubsection{\ion{H}{1} Aperture Synthesis Interferometry}

Aperture-synthesis radio observations at 21 cm have been obtained to
determine \ion{H}{1} velocity fields, from which extended rotation
curves can be derived to supplement the H$\alpha$ kinematic data, and
to map the surface density of the cold neutral gas.  In total, 43
galaxies have been imaged in \ion{H}{1} with either the VLA in its
C-short configuration (7 galaxies in 2005), the WSRT (20 galaxies in
2007-2009, 3 overlapping with VLA), or the GMRT (19 galaxies in
2008-2009, 1 overlapping with VLA and WSRT, 2 others overlapping with
WSRT). Integration times were typically 5 to 12 hours per source, with
minimum synthesized beams between 5 arcsec for the GMRT and
15$\times$30 arcsec for lower-declination sources observed with the
WSRT.  After Hanning smoothing, the FWHM velocity resolutions are
10.3, 4.1 and 13.2 km s$^{-1}$ for the VLA, WSRT and GMRT
respectively.  The column-density sensitivities of the data, smoothed
to $\sim$15 arcsec angular resolution and $\sim$12 km s$^{-1}$
velocity resolution, is 2 to 5$\times10^{20}$ atoms cm$^{-2}$ at the
5$\sigma$ level.  The observed galaxies are indicated in Table~3 with
V, W, or G, respectively, according to the facility used.  All 43
galaxies with high-quality H$\alpha$ and stellar kinematic IFU data,
and all 41 galaxies with Spitzer imaging data thus have 21-cm
aperture-synthesis observations available (Martinsson et al. 2010, in
preparation).

\subsection{Disk Mass Surface-Density}

The kinematic data critical to dynamical estimation of the disk mass
surface-density ($\sigma_z$) comes from the two-dimensional velocity-
and velocity-dispersion fields measured in the \ion{Mg}{1b} and
\ion{Ca}{2} regions. Below, we summarize the primary galaxy and stellar
template observations.

\subsubsection{\ion{Mg}{1b} and \ion{Ca}{2} integral-field spectroscopy}

Stellar kinematic data were collected with both the SparsePak and PPak
IFUs. SparsePak stellar line-of-sight velocity dispersion
($\slos$) observations were taken over 17 runs totaling 58
nights, plus portions of 4 other runs totaling 11.5 nights.  Six of
these runs were used to gather \ion{Ca}{2} data, mostly during the
pilot program from May 2001 through May 2002. SparsePak \ion{Mg}{1b}
data were gathered from April 2005 to April 2007. PPak stellar
$\slos$ observations were taken in \ion{Mg}{1b} only, over 12
runs totaling 48 nights from March 2003 to January 2007. Combined with
the Phase-A H$\alpha$ observations, the spectroscopic campaign used
$\sim$160 nights of 4m-class telescope time over 6 years.

Of the 46 galaxies observed with stellar IFS, 19 were observed in the
\ion{Mg}{1b}-region with both SparsPak and PPak (10 of which have
high-quality data from both instruments). This overlap serves to
cross-check $\slos$ derived from different instruments, and
enables us to assess the impact of instrumental artifacts. In these
cases, the data can also be combined to increase the depth.  Of the 9
galaxies with \ion{Ca}{2}-region IFS, 5 are of high-quality, and the
additional 4 galaxies have high-quality \ion{Mg}{1b}-region IFS.

Relevant science data include deep, multi-exposure, single-position
galaxy pointings for \ion{Mg}{1b} and \ion{Ca}{2}-triplet spectral
regions and short template-star observations (below). In contrast to
the H$\alpha$ observations, we did not attempt to create a filled-in
map for the galaxy observations since wherever possible we hoped to be
able to combine the multiple exposures and maximize S/N in individual
fibers at a single position. Spatial registration of these data use
the same methods employed with the H$\alpha$ data (see Paper II).
Given the limited spectral range wavelength of our configurations,
quartz-lamp dome-flats served to provide a relative flux calibration
to the spectra (to within a few percent), as verified by observations
of spectrophotometric standards.  This is adequate for preserving the
shape of the spectral continuum in both template and galaxy data,
useful for template fitting in direct-wavelength or cross-correlation
approaches.

Basic spectroscopic data processing was similar to that used for the
H$\alpha$ IFS observations. Modifications were made to implement
flexure-corrections in the PPak data (Martinsson et al., in
perparation; PMAS is a Cassegrain-mounted spectrograph), and to
improve sky-subtraction (see Bershady et al. 2005, Paper II, and
future papers in this series).
Considerable attention was paid to estimating errors in the extracted
spectra by carefully accounting for the spectral trace and the known
detector properties of read-noise and gain. Corrections for
scattered-light in the spectrograph optics have not been made to the
data, although they are appreciable in the \ion{Ca}{2} data (see
Bershady et al. 2005). The two-dimensional nature of the scattering
effect in SparsePak data is to add a small, featureless continuum to
the observed spectra, which should have no impact on the derived
velocity information.

\subsubsection{Stellar templates}

Roughly 150 template stars were observed in the same instrumental
configurations as the source observations for \ion{Mg}{1b} and
\ion{Ca}{2} regions, except stars were intentionally drifted across
roughly a dozen fibers in the course of an exposure.  This procedure
allows us to sample a range of fibers to test for effects of varying
instrumental resolution (see Paper II), and also to illuminate the
fibers in a fashion more similar to the near uniform illumination
during target exposures of galaxies (see Bershady et al. 2005).

The template library began with roughly 3-dozen stars with a wide
range of T (3000-20,000 K) and surface-gravity (luminosity classes
I-V), with some emphasis on giants (class III) in the late-F through
late-K spectral range, taken during SparsePak commissioning in May and
June 2001. Subsequent observations of 47 stars with SparsePak have
extended our template library to include more metal-poor stars,
particularly giants, expanding the dynamic range in metallicity from
-2.1$<$[Fe/H]$<$0.3.  With PPak, 72 stars were also observed, 13 hot
stars (B-F, luminosity class I-V) with -0.75$<$[Fe/H]$<$+0.15, and 59
cool stars (G-M, luminosity class III or IV) with -1$<$[Fe/H]$<$+0.35.
These include repeat observations to established estimates of
systematic errors due to subtle changes in instrumental configuration
(e.g., focus). A subsample of giant stars in our library most relevant
for our cross-correlation analysis in Paper II are shown in Figure 15
for both \ion{Mg}{1b} and \ion{Ca}{2} regions. Note that by mid-F and
hotter the line-strengths become appreciably weaker, while in the
\ion{Ca}{2} region the Paschen-series begins to dominate. At cooler
temperatures both the \ion{Mg}{1b} and \ion{Ca}{2} regions display
strong TiO band-heads, appearing prominently in the \ion{Mg}{1b}
region starting at K5, and in the \ion{Ca}{2} region at M3.

\subsection{Star-formation, Stellar Populations, and the Interstellar Medium}

We summarize the optical--NIR imaging and Spitzer observations taken
to characterize stellar emission and the dusty interstellar
medium. Combined with our IFS and \ion{H}{1} observations, these
provide estimates of the bolometric star-formation rate, the
energetics of the ISM, and a stellar populations inventory -- all a
framework for interpretation of our dynamical $\mls$ estimates.

\subsubsection{Optical and NIR Imaging}

Deep $UBI$ and shallow $VR$-band imaging data were obtained using the
KPNO 2.1m telescope for nearly all galaxies with H$\alpha$ IFS data,
and for the entire Phase-B sample. Typical exposure times were 600 sec
in $U$, 300 sec in $B$, 60 sec in $V$ and $R$, and 1200 sec in
$I$. Seeing ranged from 0.9 to 1.8 arcsec FWHM, with 1.4 arcsec a
typical value. The optical data were acquired using the T2KA CCD
detector (2048$^2$, $10.2\times10.2$ arcmin field, 0.305 arcsec
pixel$^{-1}$) over 4 runs totaling 26 nights. $B$-band images are
illustrated in Figure 9.

Deep $JHK$-band images were obtained for two thirds of the galaxies
with H$\alpha$ IFS data, again including all of the Phase-B
sample. These data were acquired with the same telescope and the SQIID
quad-channel imager (four 1024$^2$ InSb arrays, each with
$440\times460$ pixels active, $5.07\times5.28$ arcmin field, 0.69 arcsec
pixel$^{-1}$) over 6 runs totaling 35 nights. Observations were taken by
dithering the telecsope in a 3 by 3 grid so that the galaxy image
would fall at different locations within the unvignetted field of
view. This dither sequence was repeated three times for each galaxy,
slightly offsetting between cycles. Typical exposure times were 2160
sec, but the effective exposure time in the combinned dithered data
decrease towards the edges. Seeing ranged from 0.8 to 1.5 arcsec FWHM,
with 1.2 arcsec a typical value.

\subsubsection{Spitzer Space Telescope Imaging}

Measurements were designed to be efficient, complementing and
utilizing the more detailed Spitzer measurements of, e.g., the SINGS
Legacy program.  We restricted observations to 4.5, 8, 24,
and 70 $\mu$m imaging with IRAC and MIPS. Spitzer is a 0.85m
telescope, diffraction limited at wavelengths longward of 5.5 $\mu$m.
Detector pixels critically sample the diffraction core, except at 4.5
$\mu$m, where the data is undersampled.  The 4.5 and 8 $\mu$m IRAC
flux maps, sampled at $\sim$1.2 arcsec pixel$^{-1}$, combined with
$JHK$ ground-based surface-photometry are intended to allow us to
disentangle stellar photospheric emission from PAH contributions. The
8, 24, and 70 $\mu$m fluxes, sampled respectively at $\sim$1.2, 2.5,
and 10 arcsec pixel$^{-1}$, aim to allow us to characterize the
amplitude and temperature of the warm dust. The correlation of 24
$\mu$m fluxes to star-formation rates can be compared directly with
our H$\alpha$ and \ion{H}{1} measurements in matched apertures.

Our depth was targeted to give reliable surface-photometry out to
three optical disk scale-lengths, matching the extent of the kinematic
measurements constraining total disk mass-densities.  Exposures aimed
to reach S/N of 3 per spatial resolution element at 3 $h_R$, a depth
comparable to SINGS. Based on Dale et al. (2005), we estimate average
flux levels to be $0.025\pm0.005$ MJy sr$^{-1}$, $0.14\pm0.04$ MJy
sr$^{-1}$, $0.15\pm0.03$ MJy sr$^{-1}$ and $1.4\pm0.4$ MJy sr$^{-1}$
at 4.5, 8, 24. and 70 $\mu$m, respectively. Exposures were 360 sec for
IRAC 4.5 and 8 $\mu$m, 333 sec for MIPS 24 and 75 sec for MIPS 70
$\mu$m measurements, totaling 26.3 hours of telescope time for 41
galaxies. Images at 8 and 24 $\mu$m are illustrated in Figure 9.

Of the galaxies in our DMS Spitzer sample -- a subset of the Phase-B
sample described in \S5.1 -- all of them have high-quality H$\alpha$
spectroscopic and \ion{H}{1} imaging observations, all but one have
high-quality \ion{Mg}{1b}-region stellar absorption-line spectroscopy.

\section{METHODS OF ANALYSIS}

The many components of the DMS analysis are described in future papers
in this series, but two methods stand out: kinematic inclinations for
nearly face-on galaxies, and stellar velocity dispersions at low
surface-brightness for composite stellar populations. Because of their
central importance, these are outlined here.

\subsection{Inclinations}

\subsubsection{Kinematic Inclinations and Position Angles}

Our technique for determining kinematic inclination and position angle
(PA) employs fitting a single, tilted disk to all the H$\alpha$
velocities\footnote{We have also used other optical ionic tracers, but
  find H$\alpha$ to provide the best S/N and smoothest velocity
  field. For the galaxy sample with \ion{H}{1} maps, we will also use
  these data to derive independent kinematic inclinations.} across the
face of each galaxy, i.e., one zone, and a functional form for the
trend of rotation speed versus radius. In some cases, where there is
clearly a strong bar or oval distortion in the inner region, a
two-zone model is used. The detailed functional form of the rotation
curve is unimportant to the derivation of PA and kinematic inclination
so long as it is sufficiently flexible to account for the observed
range of radial shapes. The technique, developed by Andersen (2001),
and used with similar data in Andersen et al. (2001, 2003), is similar
to tilted-ring fitting. The advantage of limiting the number of radial
zones is to provide enough data points to precisely constrain a model
parameterization of the rotation curve.

To apply these values to the stellar data we make the reasonable
assumptions that the gas and stars are co-planar and share the same
barycenter. A detailed description of the method as applied to the DMS
will be presented in Andersen et al. (2010, in preparation).
Preliminary estimates of the performance in terms of inclination
random error are given in Paper II.  The zero-point and slope of the
nearly face-on TF relation based on kinematic inclinations using this
method (Andersen \& Bershady 2003) is consistent within the
uncertainties for the same quantities for more inclined samples,
demonstrating there are no significant systematic errors.

\subsubsection{Inclinations from Scaling-Relation Inversion}

Disk inclination can also be estimated by inverting the TF relation
(Rix \& Zaritsky 1995), assuming survey galaxies lie along a fiducial
relation established with galaxies of similar type that have reliable
kinematic or photometric inclinations.  So-called inverse Tully-Fisher
(iTF) inclinations have the advantage that their errors decrease with
inclination. Inclination from iTF are only effective if a precise
measure of the projected rotation velocity can be obtained. This
requires measuring a rotation curve rather than an integrated
line-width in order to (a) avoid confusion between rotation and
turbulent motion, and (b) verify the regularity of the velocity field
and its asymptotic flatness at larger radii (see discussion in
Verheijen 2001). The smaller the scatter in the TF relation, the more
precise the iTF-based inclination estimates.

The combination of high spectral-resolution IFS and near-infrared
photometry allows us to make the best possible inclinations estimates
from iTF. As Figure 16 illustrates for a very-nearly face-on galaxy
in our survey, whatever the true inclination and circular speed, our
spectroscopy has the resolution and spatial coverage to determine that
this galaxy has a very regular velocity field and flat asymptotic
rotation curve. The smallest observed scatter in the TF relation
requires near-infrared luminosities and verification of
asymptotically-flat, regular velocity fields (Verheijen 2001).

Despite this promise of high precision, using iTF remains a
potentially fatal path if used in isolation because its accuracy is
not verifiable a priori.  Our preference is to work in a regime where
kinematic inclination errors are small enough to verify a galaxy's
location on the Tully-Fisher relation. We will, however, take
advantage of the iTF method to reduce errors in some situations,
depending on the {\it observed} regularity of the ionized-gas velocity
fields.

\subsection{Stellar Velocity Dispersions}

To arrive at the kinematic line-width required to estimate the disk
mass surface-density, two steps are needed.  First $\slos$ must
be derived from the spectroscopic data, and second $\sigma_z$ must be
deprojected from the former line-of-sight quantity.

There is a considerable variety of methods for determining
$\slos$ from the integrated star-light of galaxies, but the
methods basically fall into the two categories of direct-fitting and
cross-correlation. In the former, galaxy spectra are fit directly in
wavelength space, with a template that is optimized for suitability
both in terms of velocity broadening and spectral match.  We adopt,
instead, a new application of the cross-correlation approach (Paper
III; Westfall 2009). As we demonstrate in Papers II and III, our
method not only optimizes the determination of the primary measure of
interest (broadening), but allows us to minimize template mismatch as
well as, if not better than, direct-fitting methods.

Within this cross-correlation context, we have considered two
complementary approaches to deriving $\sigma_z$ from our line-of-sight
measurements, each taking different advantages of the bi-dimensional
nature of our spectral coverage. Specifically, these approaches either
minimize systematics due to S/N and template-mismatch associated with
determining $\slos$, or mimize systematics associated with
deprojecting the SVE to find $\sigma_z$.

The first approach is to average fibers at a given radius over azimuth
(i.e., in a ring), correcting for their velocity offsets due to
projection of the internal velocity field, derive the broadening
function $\slos$ for the registered and combined spectrum at each
radius, and correct for line-of-sight projection of $\sigma_R$,
$\sigma_\theta$, beam-smearing, and instrumental resolution.

The number of fibers in an annulus range from 6-30 for SparsePak when
the radial width of the fiber-centers is constrained to be of order
the fiber diameter (Figure 4). Comparable-width annuli for PPak
include up to 120 fibers, but thinner annuli can be defined when the
S/N is high enough. The advantage of fiber-averaging is that higher
S/N is achieved before the cross-correlation (or direct fitting)
process to derive $\slos$. The benefits are severalfold. Higher
S/N enables better, direct inspection as well as quantitative
assessment of goodness of fit, particularly in terms of template
mismatch. In contrast, only velocity information is required from
individual fibers in order to velocity-register the spectra, which is
much easier to derive at low S/N, and is rather immune to template
mismatch. We show in Paper II that this registration is quite robust
when assisted with prior empirical knowledge based on, e.g., the gas
or stellar rotation curve. Higher S/N also helps ensure that any
noise-dependent systematics in estimating $\slos$ are
minimized, although we have not found such systematics to be present
at a significant level for S/N$>3$ pix$^{-1}$ (see Papers II and III).
A further advantage is that the derived $\slos$ at any one
radius relative to another is independent of model assumptions, e.g.,
on the radial form of $\sigma_z$. The only disadvantage is that some
prior knowledge of the SVE must be known in order to decompose
$\slos$ to find $\sigma_z$.

To handle this short-coming, we have devised a second approach,
described in Westfall (2009) and Westfall et al. (2010, in preparation),
which treats the fibers separately, in that their $\slos$ is
determined individually (e.g., by cross-correlation). There is ample
signal in inner regions of most galaxies (typically out to 1 to 2 disk
scale-lengths) to examine the velocity broadening in this way. By
preserving the two-dimensional information on $\slos$, we may
solve directly for the SVE by imposing at least one other
constraint. The most powerful constraint is dynamical, and arises from
the asymmetric-drift equation. This takes full advantage of the direct
observations of the lag between the projected tangential velocity
between gas and stars, as illustrated in Figure 17 for UGC 6918. This
lag, akin to the so-called asymmetric-drift observed in the Milky Way,
is an interesting and dynamically significant measure in its own
right. Figure 17 demonstrates the repeatability of our velocity
measurements at different wavelengths (513 and 670 nm), with different
instruments (SparsePak and PPak), and different tracers (gas and
stars). Moreover, the quality of the velocity-centroid measurements
are high enough and the velocity-fields are regular enough to reveal
asymmetric-drift at levels of 5 to 10 km s$^{-1}$, well below our
instrumental resolution.

Further dynamical constraints include the epicycle approximation, and
other equations that establish a covariance between any two axes of
the SVE, e.g., constant vertical-anisotropy ($\sigma_z/\sigma_R$
constant), or constant disk stability. Radial parameterizations are
also of aid in solving for the trend in the SVE amplitude and
shape. Westfall (2009) describes the various possible methods in
detail, and demonstrates their application. The key advantage of this
approach is the direct measure of the SVE shape, which we find is
fairly robust to different dynamical assumptions and
parameterizations. Further, individual fiber measurements allow for
higher spatial resolution of $\slos$. In particular, the
azimuthal structure can be sensitive not only to the changing SVE
projection but spiral structure. However, the impact of
template-mismatch is difficult to assess given the limited S/N of
individual fiber spectra, the radial range of analysis is similarly
limited, and the parameterization of the SVE as well as the use of
dynamical assumptions has the potential to introduce systematic
errors.

On balance, the two approaches make different trades in terms of
systematic and random errors, and therefore are best used together, in
an iterative sense, respectively probing template mismatch and the
SVE shape. Paper II describes the elements critical to the
first (fiber-averaging) approach. This is useful for visualizing the
quality of the spectral data, and it is used to estimate the survey
error budget.

\section{SUMMARY}

We have argued that the disk-halo degeneracy, central to obstructing a
reliable mass decomposition based on the rotation curves of spiral
galaxies, is unbreakable without independent information about either
the halo or disk mass, or a combination of the two. Despite the
long-known limitations of using rotation curves to place limits on
$\mls$ (van Albada et al. 1985), namely that only upper limits
may be set (the maximum-disk hypothesis), the literature abounds with
efforts to use rotation curves to place stronger constraints on
$\mls$, dark-matter halo parameters, or worse -- both. The
uncertainties do not go away with better data (e.g., de Blok et
al. 2008) or with two-dimensional fitting. This is because galaxies
are intrinsically non-axisymmetric at a level (5-10 km s$^{-1}$ or
more; Trachternach 2008) required to discern either significant
differences between low and high values of $\mls$, different
halo models, or both. Unfortunately, independent contraints from SPS
models on $\mls$ are rather poor (uncertainties of factors of 4,
and possibly much larger are likely) due to uncertainties in the
low-mass end of the IMF, late-phases of stellar evolution (especially
the TP-AGB), and a wide range of possible star-formation and
chemical-enrichment histories.

We have described one of several methods in the literature that can be
used to break the disk-halo degeneracy, and hence challenge the
maximum-disk hypothesis. This method uses measurements of the vertical
velocity dispersion of disk stars along with an estimate of the
vertical scale height to place direct dynamical limits on the mass
surface-density of spiral disks. Advantages of the approach are that
it is conceptually simple, it uses collisionless tracers, and it
provides direct, spatially resolved information about the mass
distribution within galaxies. The measurement was suggested and
pioneered 25 years ago (Bahcall \& Casertano, 1984; van der Kruit \&
Freeman 1984). One small, early survey (Bottema 1993) yielded the
tantalizing result that normal galaxy disks are substantially
sub-maximal ($\sim$50\% by mass).  However, until recently there have been
substantial technical limitations in gathering sufficient signal to
measure reliably the velocity dispersions of disk stars at the low
surface-brightness levels required to probe well into the disks of
normal spirals.

In this context we have given an overview of a new study, the DiskMass
Survey, of the mass distribution within normal spirals galaxies.  Our
approach has been to use custom-built IFUs to measure the stellar and
gaseous velocities and velocity dispersion fields in nearly-face on
systems close to 30 degrees. Inclinations are low enough to favorably
project $\sigma_z$, but not too low to prevent kinematic inclinations,
and hence total mass estimates to be made based on the observed
velocity fields. The sample selection and survey protocol, whereby we
down-select targets for ever increasing study, has been detailed
here. The resulting data set includes H$\alpha$ velocity fields for
146 galaxies; stellar kinematics in at least one of the \ion{Mg}{1b}
(513 nm) or \ion{Ca}{2} (860nm) spectral regions for 46 of these
galaxies; and Spitzer Space Telescope IRAC and MIPS (4 through 70
$\mu$m) imaging as well as 21cm aperture-synthesis imaging for 41 of
these 46 galaxies. Observations required $\sim$220 nights of 2-4m
aperture ground-based telescope time, 400 hours of 21cm radio
interferometry, and 26 hours of Spitzer time. The DMS galaxies span
close to an order of magnitude in color and size (surface-brightness),
and two orders of magnitude in luminosity.  The size and scope of the
survey is large enough to make reliable statistical statements about
disk maximality and calibrate $\mls$ over a wide range of disk
properties.

We have briefly summarized the observations and analysis undertaken
for the survey to serve as a road-map to later papers in this series.
In a companion paper we present a detailed error budget for disk mass
surface-density, mass-to-light ratios and disk mass-fraction (Paper
II); a following paper establishes our new cross-correlation method
(Paper III). Future papers will present results on topics including,
but not limited to, the measurement of total galaxy mass; the
Tully-Fisher relation for nearly face-on galaxies; disk kinematic
asymmetries, star-formation rates, gas content, mass surface-density,
the SVE, and $\Upsilon$; as well as halo density profiles. While a few
of these topics will have been explored in greater depth by other
surveys (e.g., star-formation; SINGS, Kennicutt et al. 2006), our aim
is to complement and extend our knowledge of nearby galaxies by
providing an absolute calibration of stellar mass, and a direct
inference of the shape of dark-matter halos. This advance, in turn,
should extend the bounds of our understsanding of how galaxies have
formed and evolved.

\acknowledgements We thank P. C. van der Kruit for his insight and
encouragement, and R. de Jong for a helpful review of the
manuscript. M.A.B. is indebted to R. Kron and S. Majewski for their
inspiration, starting with O'Dell's Nebular Spectrograph at Yerkes
Observatory. We acknowledge contributions to this project from
A. Schechtman-Rook, Y. Tai, and K. Kern. Research was supported under
NSF/AST-9618849, NSF/AST-997078, NSF/AST-0307417, AST-0607516
(M.A.B.); Spitzer GO-30894 (RAS and MAB); NSF/OISE-0754437 (K.B.W.).
M.A.W.V. and T.M.  acknowledge the Leids Kerkhoven-Bosscha Fonds for
travel support.  M.A.B. acknowledges support from the Wisconsin Alumni
Research Foundation Vilas Fellowship and the University of Wisconsin
College of Letters \& Science Ciriacks Faculty Fellowship. This
research has made use of NED\footnote{\rm
  http://nedwww.ipac.caltech.edu/}, DSS\footnote{\rm
  http://archive.stsci.edu/dss/acknowledging.html}, and
SDSS\footnote{\rm http://www.sdss.org/collaboration/credits.html} data
archives.

\appendix

\section{Corrections for RC3 and 2MASS $B$ and $K$ Total Magnitudes}

The total magnitudes from RC3 and 2MASS, listed in NED as ``{\tt B
  (m\_B)}'' (here $m_B$) and ``{\tt K\_s\_total (2MASS)}'' (here
$K_{\rm s,total}$) have been corrected for systematic effects with
apparent magnitude based on comparison to total magnitudes measured
for the samples of Verheijen \& Sancisi (2001) and de Jong (1994). The
latter are based on deep surface-photometry. These corrections,
illustrated in Figure 18, are:
\begin{equation}
B \ = \ m_B \ - \ 6.18 \ + \ 1.08 \ m_B \ - \ 0.046 \ m_B^2
\end{equation}
\begin{equation}
K \ = \ K_{\rm s,total} \ - \ 5.73 \ + \ 1.30 \ K_{\rm s,total} - 0.074 \ K_{\rm s,total}^2
\end{equation}
and are applied to the photometry listed in Table 2.  We emphasize
that these corrections, along with the magnitudes in Table 2 are
corrected neither for internal or foreground extinction, or
$k$-corrections, and are applicable to extended sources.

Noordermeer \& Verheijen 2007 find a surface-brightness dependence to
their comparison of the Verheijen \& Sancisi (2001) data, but we see
no such additional dependence after applying the above
corrections. We note that in the Verheijen \& Sancisi (2001) sample,
which is volume limited, apparent magnitude and surface-brightness are
highly correlated.

The dispersions in the corrected values are 0.31 mag for $B$, compared
to the mean expected random error of 0.25 mag (0.22 mag for $m_B$ and
0.08 mag for the reference data).  For the $K$ band there is 0.36 mag
dispersion in the corrected values, compared to the mean expected
random error of 0.11 mag (0.05 mag for $K_{\rm s,total}$, and 0.09 mag for the
reference data). However, the dispersion is dominated by faint sources
where the expected random errors are larger.  The dispersion in the
corrected $B-K$ is 0.30 mag, compared to the mean expected random
error of 0.24 mag. 

We conclude that our corrections for total galaxy magnitudes can be
safely applied to $m_B=15.5$ and $K_{\rm s,total}=12$ ($B=15$ and $K=11.5$ in
Table 2). Corrections to $K_{\rm s,total}=13.5$ ($K=12$) are likely accurate
in the mean, but with reduced precision.  For the $B$ band this
includes all but one source (UGC 3965), but this source is a BLAGN so
the interpretation of its photometry is suspect in any event.  For the
$K$ band, 11 sources are fainter than $K=11.5$, but only one (UGC
6616) is fainter than $K=12$.

\clearpage

\begin{center}

                                 
\clearpage

\begin{figure}
\figurenum{1}
\epsscale{0.8}
\plotone{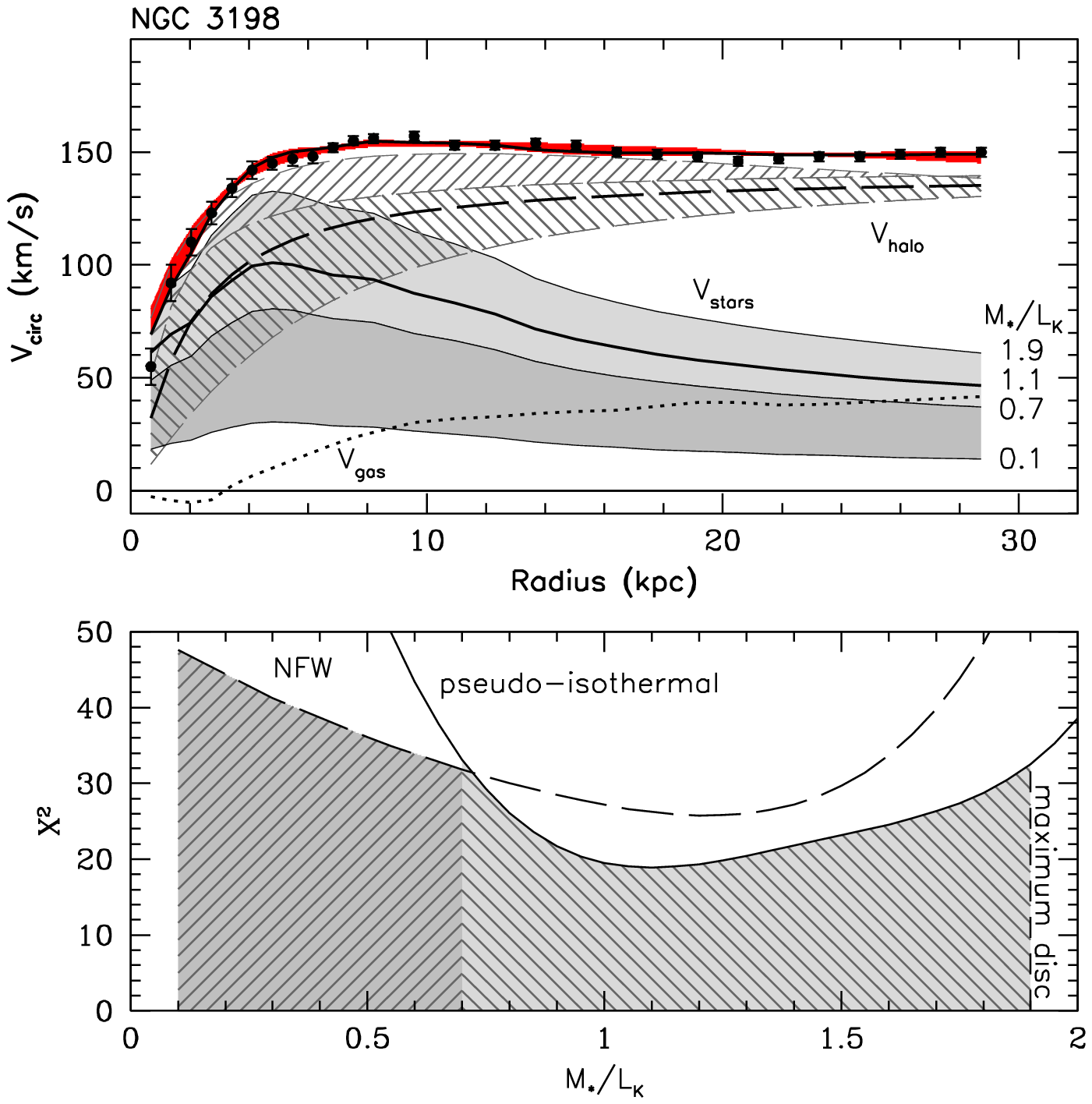}
\caption{Example mass-decompositions for NGC 3198 for a factor of
  $\sim$20 range in $K$-band $\mls$ (assumed constant with radius),
  constrained by the observed 21 cm (\ion{H}{1}) rotation curve
  (points) and gas mass surface-density, adopting either a spherical
  pseudo-isothermal or NFW dark-matter halo profile.  Stellar and halo
  mass-ranges (upper panel) are grey-scale coded (medium and light)
  for the break-point between halo models seen in the formal $\chi^2$
  in the bottom panel. Total mass for the full range of $\mls$ is
  shown (upper panel) in red (dark grey).  All $\mls$ values provide
  indistinguishably good fits to the observed rotation curve given the
  amplitude of non-axisymmetric motion (see text). The disk radial
  scale-length is 2.7 kpc. [COLOR IN ELECTRONIC EDITION ONLY]}
\end{figure}

\clearpage

\begin{figure}
\figurenum{2} 
\epsscale{0.9} 
\plotone{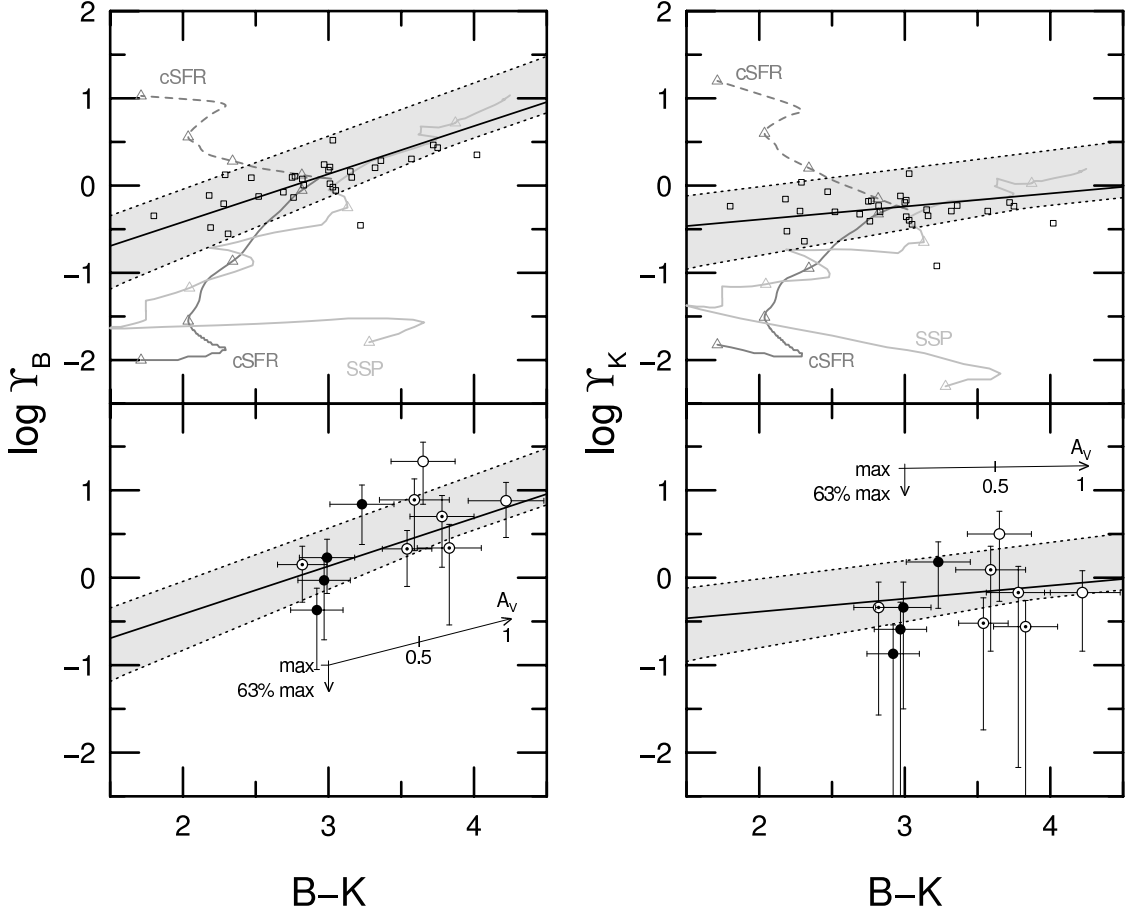}
\caption{$B$- and $K$-band $\Upsilon$ versus $B-K$ color. {\it
    Models:} Solid black line and gray areas represent mean and
  extrema $\mls$ for the full range of SPS models given in Bell
  \& de Jong (2001); they scale their IMF such that no galaxies in the
  Verheijen (1997) sample are super-maximal (see text). Trends in
  $\mls$ for SSP and cSFR models are plotted in the top panels
  as light and medium-gray lines punctuated by triangles marking age
  in time-steps from 10$^7$ to 10$^{10}$ years in equal dex intervals
  (Bruzual \& Charlot 2003). For cSFR, $\mldyndisk$ is also
  calculated including gas mass in future star-formation up to
  10$^{10}$ years (dashed line).  {\it Observations:} $\mldyndisk$
  values and uncertainties for galaxy disks in the literature (see
  text).  Open squares are estimates (Sanders \& Verheijen 1998; top
  panel) based on rotation-curve decomposition assuming MOND for the
  same sample of galaxies later used by Bell \& de Jong (2001).
  Circles (bottom panel) represent the Bottema (1993) sample, based on
  direct kinematic measurement marked here by inclination for face-on
  ($i<30^\circ$; filled), highly-inclined ($70^\circ<i<80^\circ$;
  dotted), and edge-on ($i>80^\circ$; open) systems.  Reddening
  vectors are indicated, marked for 0.5 and 1 mag of visual extinction
  ($A_V$) in a foreground screen. Change in $\mls$ from a
  so-called maximum disk (providing up to 85-90\% of the observed
  galaxy rotation) to a disk which provides only 63\% of this rotation
  is also indicated.}
\end{figure}

\clearpage

\begin{figure}
\figurenum{3}
\epsscale{0.9}
\plotone{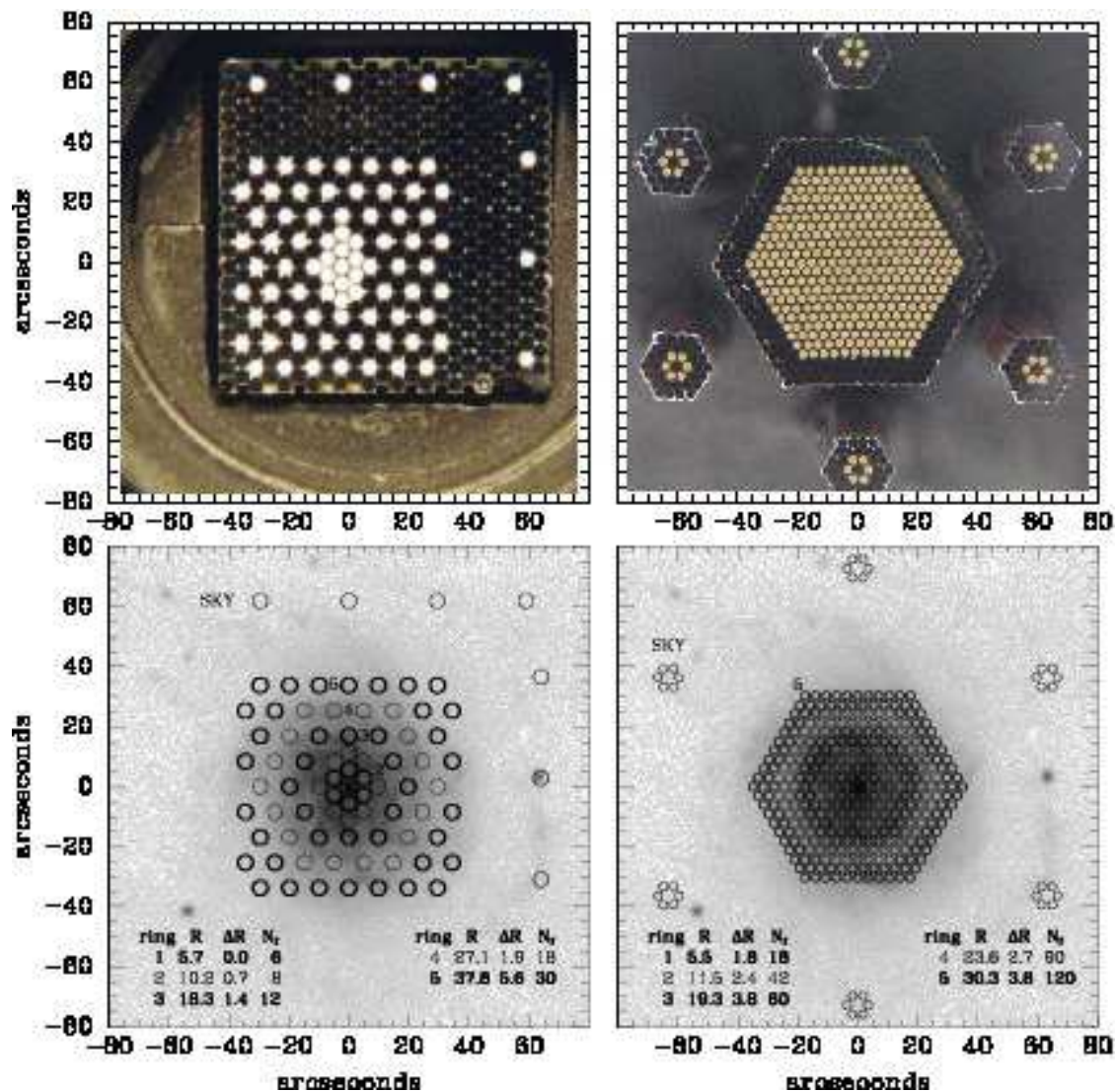}
\caption{(Top row) Fiber integral-field units (IFU) developed for the
  DiskMass Survey, shown at the same angular scale: SparsePak (left,
  WIYN 3.5m telescope) and PPak (right, Calar Alto 3.5m
  telescope). Both IFUs subtend approximately 70 arcsec field in their
  core, sample with 4.7 and 2.7 arcsec diameter fibers, respectively,
  and reformat to feed grating-dispersed spectrographs in long-slit
  mode.  Relevant spectrograph configurations, achieving
  $8000<\lambda/\delta\lambda<11000$ are given in Table 1.  (Bottom
  row) SparsePak and PPak overlays on UGC 6918 $B$-band image from the
  KPNO 2.1m telescope, with examples of azimuthal rings of fibers
  identified. Inset tables specify mean sky-plane radius ($r$),
  half-range of radius ($\Delta r$) for fiber centers, and number of
  fibers per ring (N$_f$).}
\end{figure}

\clearpage

\begin{figure}
\figurenum{4}
\epsscale{1.0}
\plotone{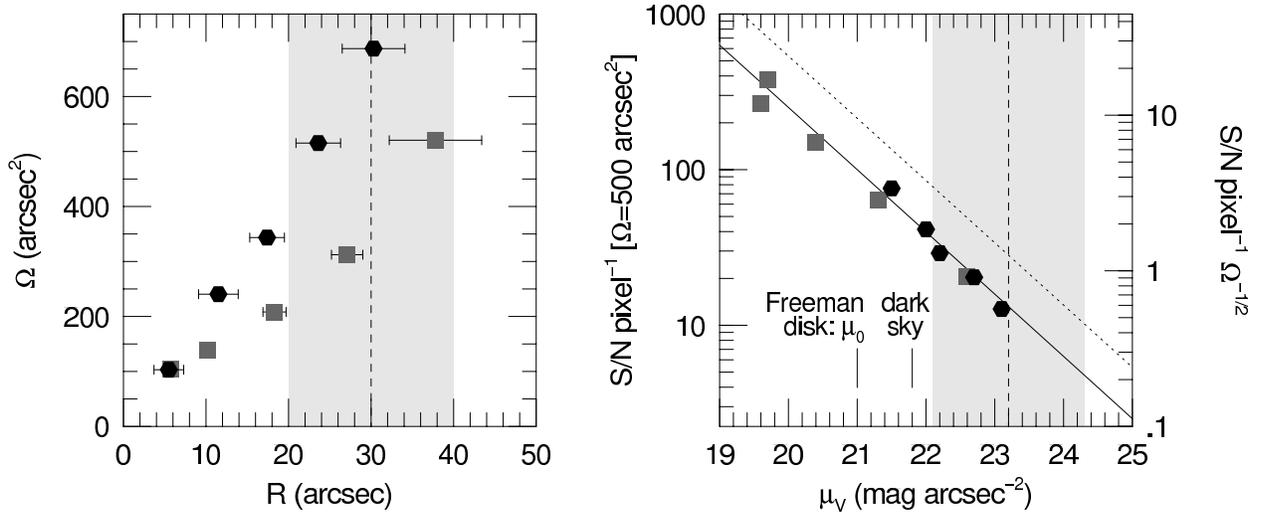}
\caption{(Left panel) Solid-angle versus radius in five radial bins
  for SparsePak (grey squares) and PPak (black hexagons). Horizontal
  bars represent radial range in fiber centers for each bin. (Right
  panel) S/N per pixel as a function of $V$-band surface-brightness
  for stellar absorption-line spectroscopy in the \ion{Mg}{1b} region in a
  ring subtending 500 arcsec$^2$. Rightmost vertical scale gives
  equivalent S/N per arcsec$^2$. Data for UGC 6918 (SparsePak) and UGC
  1635 (PPak) are scaled to this $\Omega$ and typical total
  integration-times with these instruments (12 and 8 hr,
  respectively).  Diagonal lines are the predicted, detector-limited
  performance for the typical (solid) and longest (dotted) total
  integration times. In both panels grey shaded regions in both panels
  mark the distribution of apparent size and surface-brightness at
  $2.25 h_R$ encompassing 90\% of the galaxies in the DiskMass
  sample; vertical dashed lines represent median values.}
\end{figure}

\clearpage

\begin{figure}
\figurenum{5}
\epsscale{1.0}
\plotone{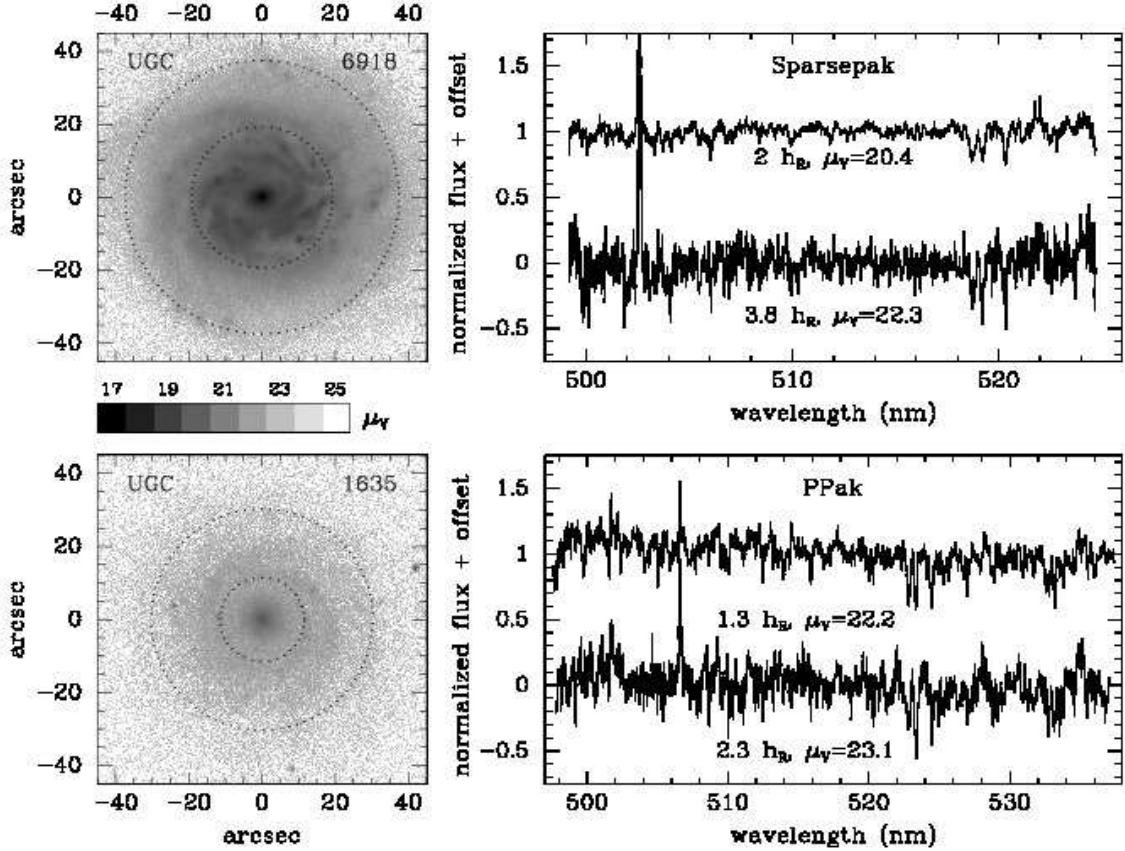}
\caption{Survey spectroscopic dynamic range in surface-brightness,
  shown with KPNO 2.1m $V$-band images (60 sec exposures) and spectra
  in the \ion{Mg}{1b} region for UGC 6918 (top; 2.25 hours SparsePak
  total integration in 3 exposures) and UGC 1635 (bottom; 6 hours PPak
  total integration in 6 exposures). These galaxies span the
  surface-brightness range of our survey. Image grey-scale is
  logarithmic, calibrated in magnitudes. Radial bins, labeled on
  spectra and marked on images, consist of 12 and 18 fibers
  (subtending 208 and 382 arcsec$^2$ solid-angle) for SparsePak, and
  60 and 120 fibers (subtending 344 and 687 arcsec$^2$ solid-angle)
  for PPak. Spectra are plotted in the observed frame of each
  instrument in normalized flux units. Mean S/N per pixel is 42 and
  10.5 for SparsePak spectra of UGC 6918 at 2 and 3.8 h$_R$, and 21
  and 13 for PPak spectra of UGC 1635 at 1.3 and 2.3 h$_R$.}
\end{figure}

\clearpage

\begin{figure}
\figurenum{6} 
\epsscale{0.9} 
\plotone{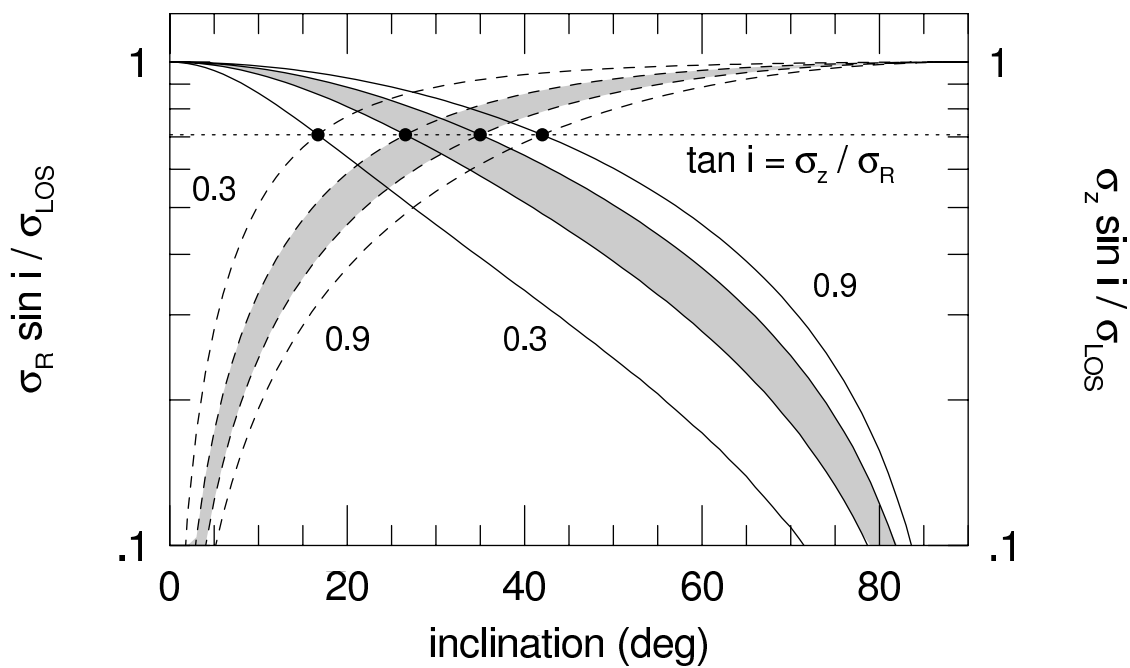}
\caption{Projection of vertical and radial disk velocity-dispersion
  components ($\sigma_z$, solid lines; $\sigma_R$, dashed line) into
  the minor-axis line-of-sight as a function of inclination.  Curves
  represent different assumed SVE shape parameterized by the ratio
  $\sigma_z / \sigma_R$. Ratios of 0.5 to 0.7, expected to be typical
  for galaxies in our survey (Paper II), are high-lighted in
  grey. Horizontal, dotted-line intersections with curves mark
  inclinations where radial and vertical projections are equal.
  Inclination values (i) at these intersections are related to the SVE
  axis ratio, as given in the figure.}
\end{figure}

\clearpage

\begin{figure}
\figurenum{7}
\epsscale{1.0}
\plotone{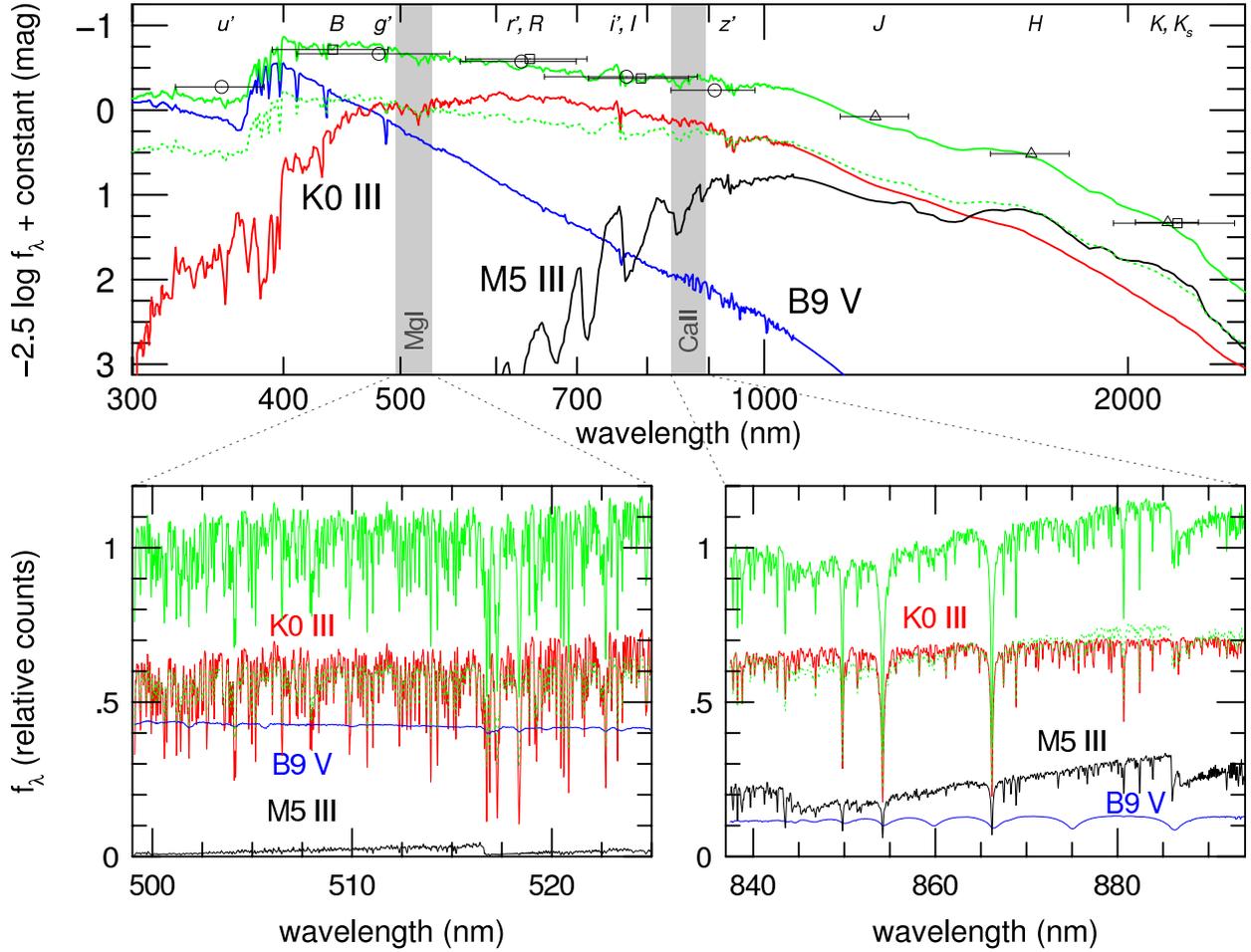}
\caption{Three-star spectral-synthesis for UGC 6918, with integrated
  colors typical of galaxies in the DMS, at low-resolution (top panel)
  and high-resolution (bottom panels).  Low-resolution spectra are
  from the compilation of Bruzual \& Charlot (1993). High-resolution
  spectra are from our own echelle stellar-template library observed
  with SparsePak, and show the spectral regions used to measure
  $\slos$ in the DMS. Constraints on the spectral mix, described in
  the text, are based on broad-band photometry shown here as squares
  [$BVIK$ magnitudes from Verheijen (1997)], circles ($u'g'r'i'z'$
  magnitudes based on our own photometric measurements on SDSS
  images), and triangles ($JHK_{\rm s}$ measurements from 2MASS
  corrected as described in Appendix A).  Individual star spectra are
  labeled (blue, red, black for B9~V, K0~III, and M5~III,
  respectively). In the \ion{Mg}{1b}-region and \ion{Ca}{2}-region the
  flux ratios B9:K0:M5 are 0.42:0.56:0.02 and 0.12:0.65:0.23,
  respectively.  Green solid curves show the combined spectrum using
  this mix. Green dotted curves are the combined spectrum renormalized
  to the dominant stellar spectrum in the \ion{Mg}{1b} region (top and
  bottom left panels) and \ion{Ca}{2} region (bottom right panel),
  drawn to indicate the difference with a one-star template. [COLOR]}
\end{figure}

\clearpage

\begin{figure}
\figurenum{8}
\epsscale{0.7}
\plotone{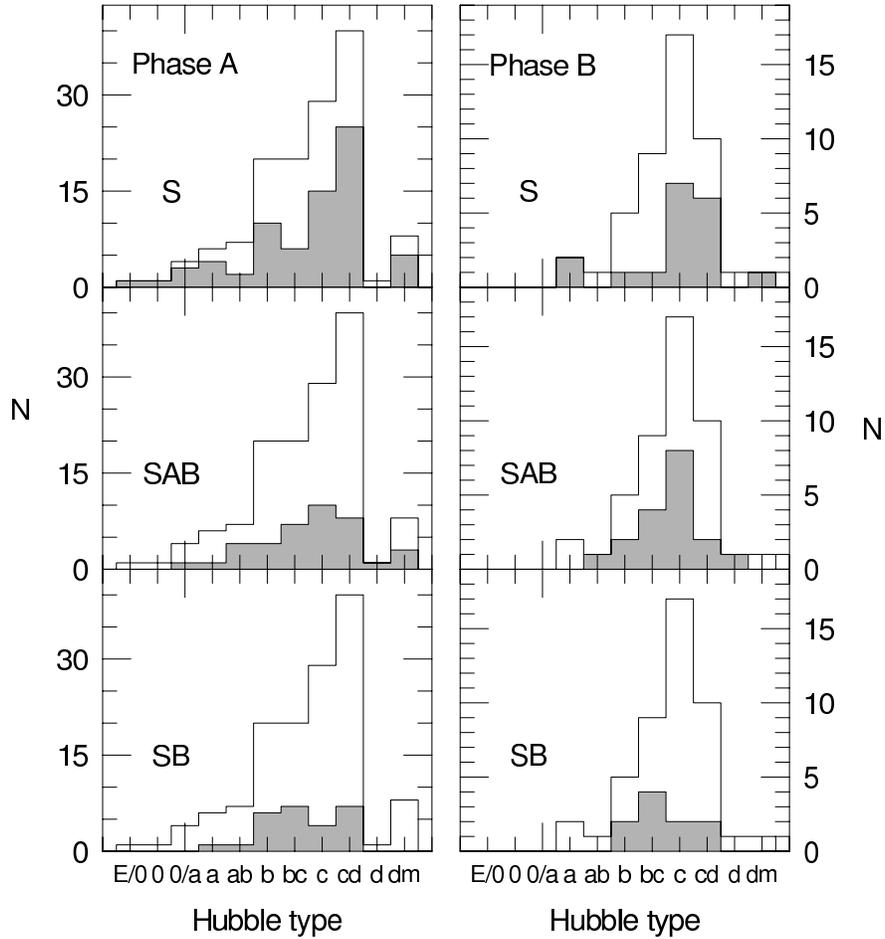}
\caption{Distribution of Hubble types in Phase-A (left) and Phase-B
  (right) samples, broken down by un-barred (S), weakly-barred (SAB),
  and barred (SB) systems (grey histograms). Open histograms represent
  the total sample for all families; these are identical for the three
  panels in each column.}
\end{figure}

\clearpage

\begin{figure}
\figurenum{9}
\epsscale{1.0}
\plotone{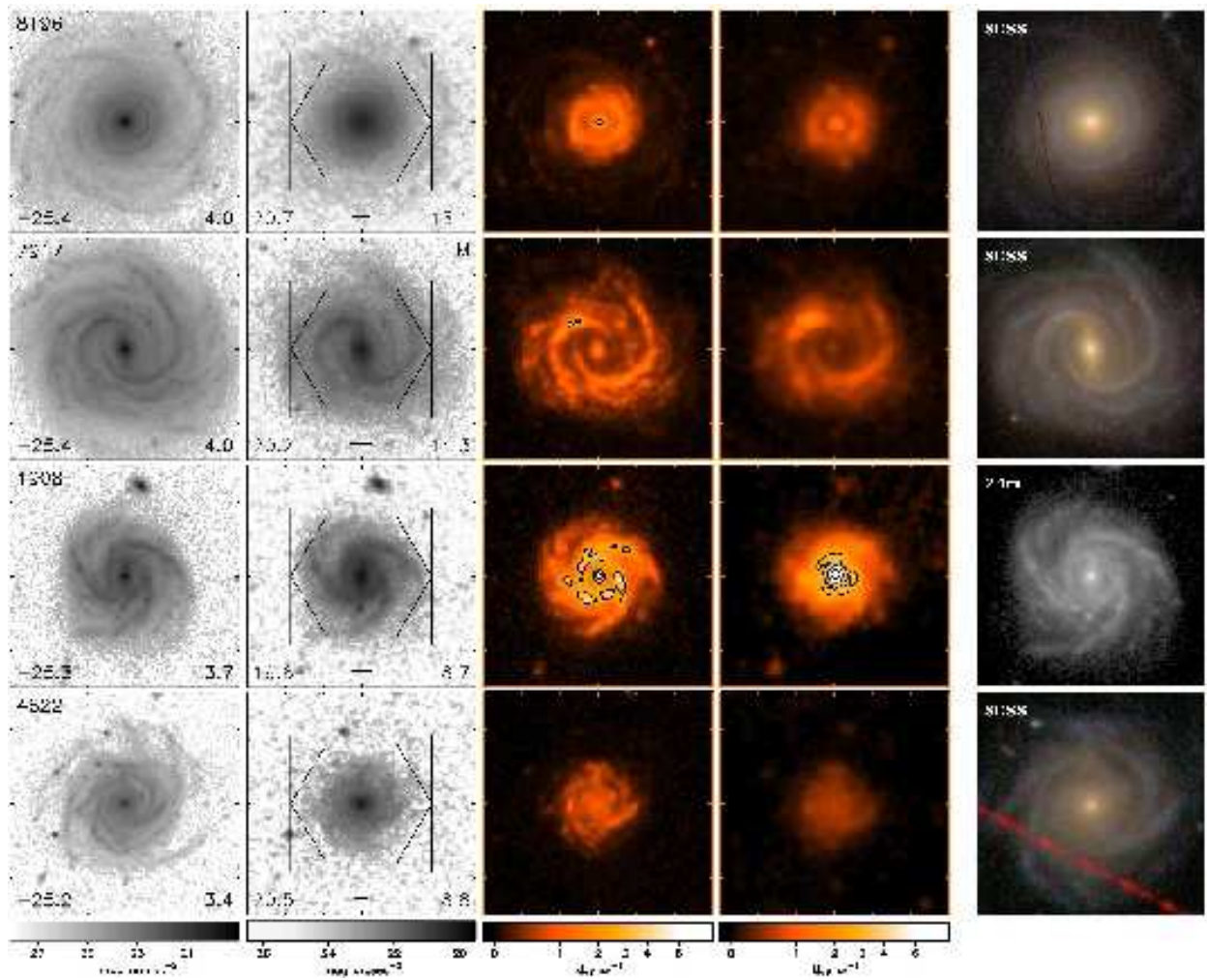}
\caption{Montage of Phase-B Spitzer sub-sample, sorted by $K$-band
  absolute magnitude, one line per galaxy. Images from left to right:
  KPNO 2.1m $B$-band, POSS-II red band used for sample selection,
  Spitzer/IRAC 8 $\mu$m, Spitzer/MIPS 24 $\mu$m, SDSS false-color
  (when available) or KPNO 2.1m $B$-band, as indicated by label.  (For
  UGC 72 neither 2.1m nor SDSS images were available; here a POSS-II
  blue image was used.) SDSS images are SkyServer g'r'i' (3-band)
  composites. Right-most images are scaled to 50$\times$50 kpc
  rasters; all other images are 2$\times$2 arcmin rasters. Gray- and
  color-scale calibration in mag arcsec$^{-2}$ or MJy sr$^{-1}$ are
  given at the bottom. Contours in 8 and 24 $\mu$m images are at 4, 8,
  16, 32, 64, and 128 MJy sr$^{-1}$. In the first column is given the
  UGC \# (upper-left), $K$-band absolute magnitude (lower-left), and
  $B-K$ color (lower-right). In the second column is given the
  $R$-band disk central surface-brightness in mag arcsec$^{-2}$
  (lower-left) and scale length in arcsec (lower-right). SparsePak
  square footprint (all H$\alpha$ observations; MgI and CaII
  observations are indicated by ``M'' and ``C'' above the upper-right
  corner of the footprint) and PPak hexagonal footprint (MgI
  observations) as well as a 5 kpc bar are drawn in the second
  column [COLOR]}.
\end{figure}

\clearpage

\begin{figure}
\figurenum{9}
\epsscale{1.0}
\plotone{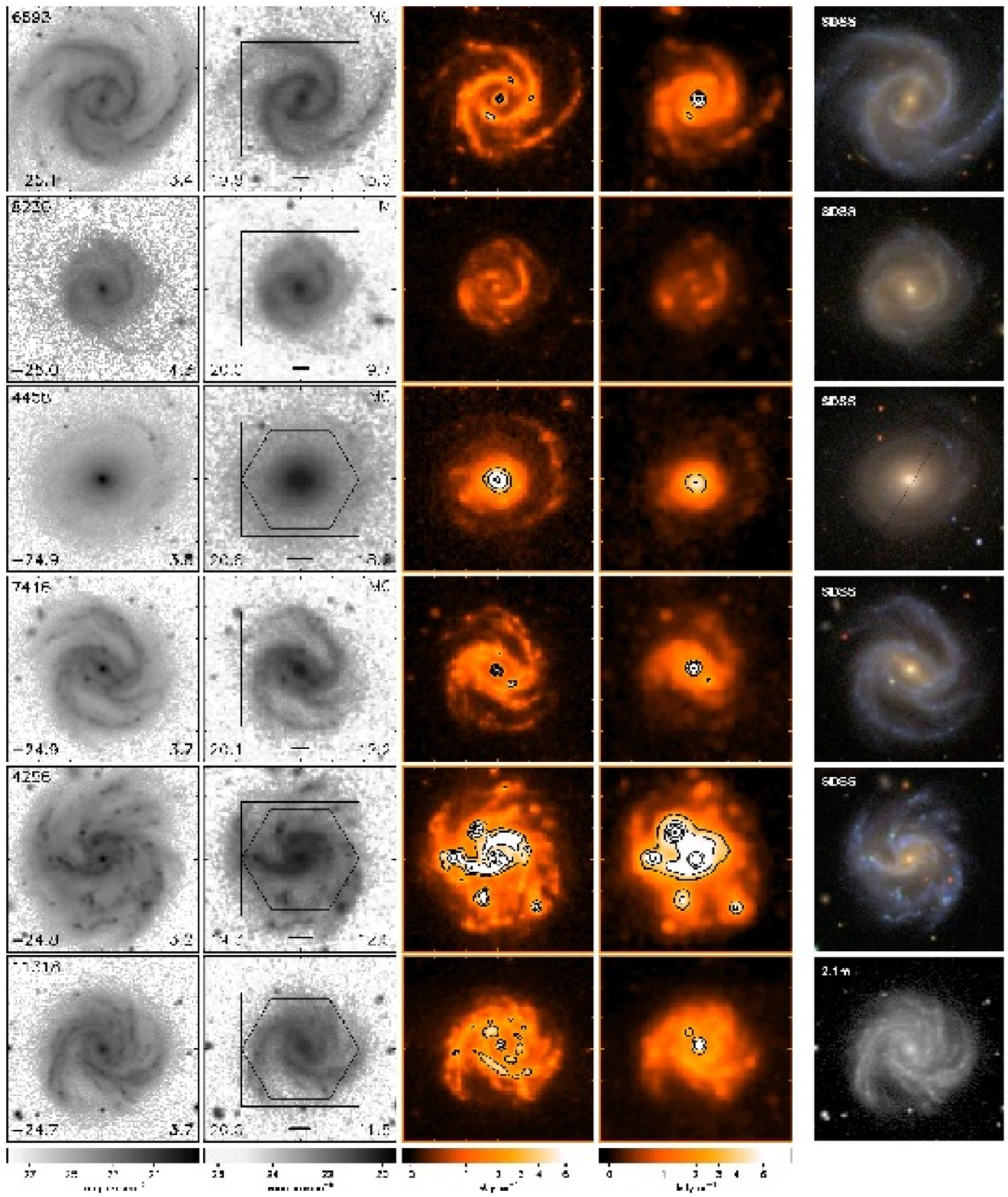}
\caption{continued [COLOR]}
\end{figure}

\clearpage

\begin{figure}
\figurenum{9}
\epsscale{1.0}
\plotone{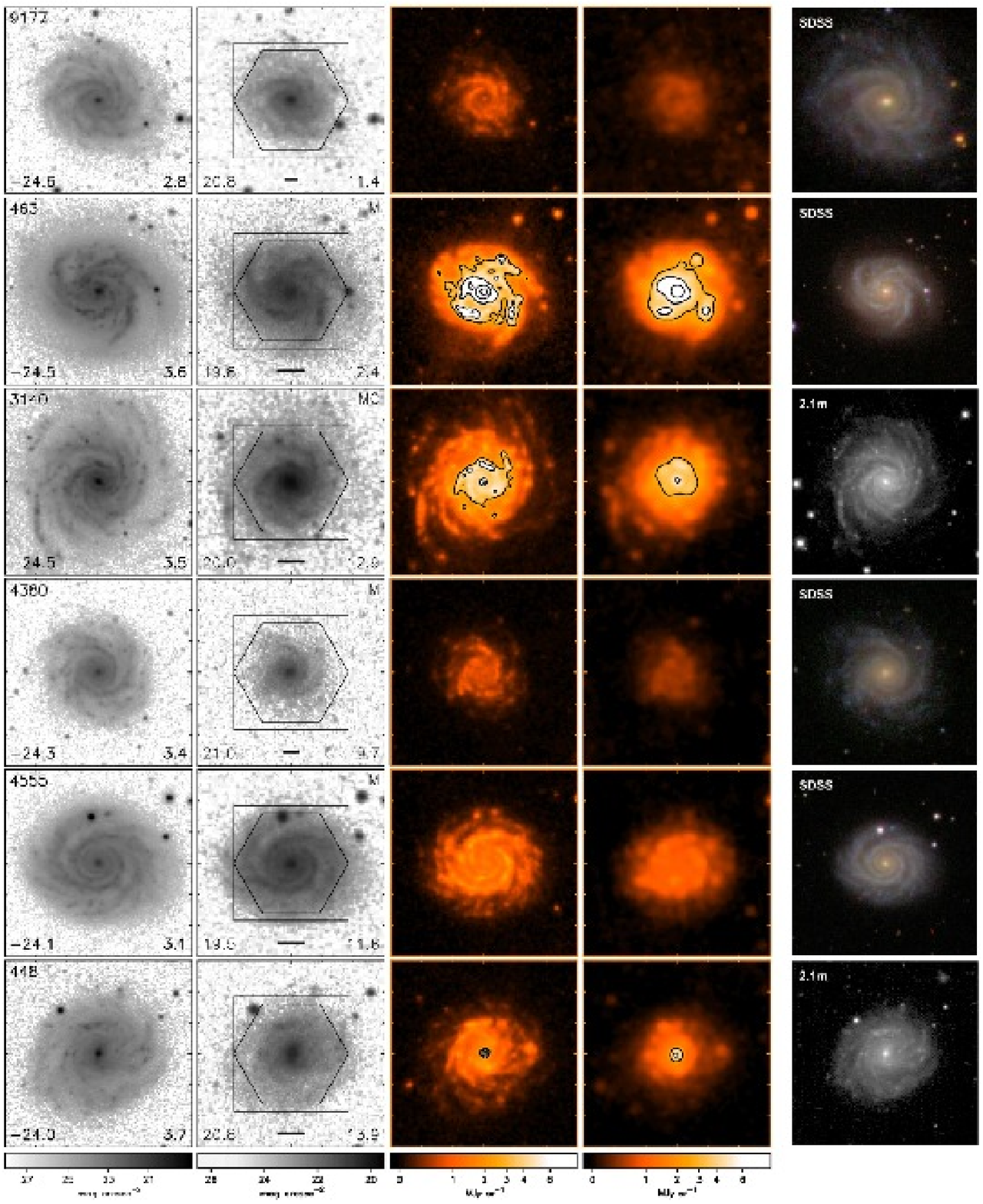}
\caption{continued [COLOR]}
\end{figure}

\clearpage

\begin{figure}
\figurenum{9}
\epsscale{1.0}
\plotone{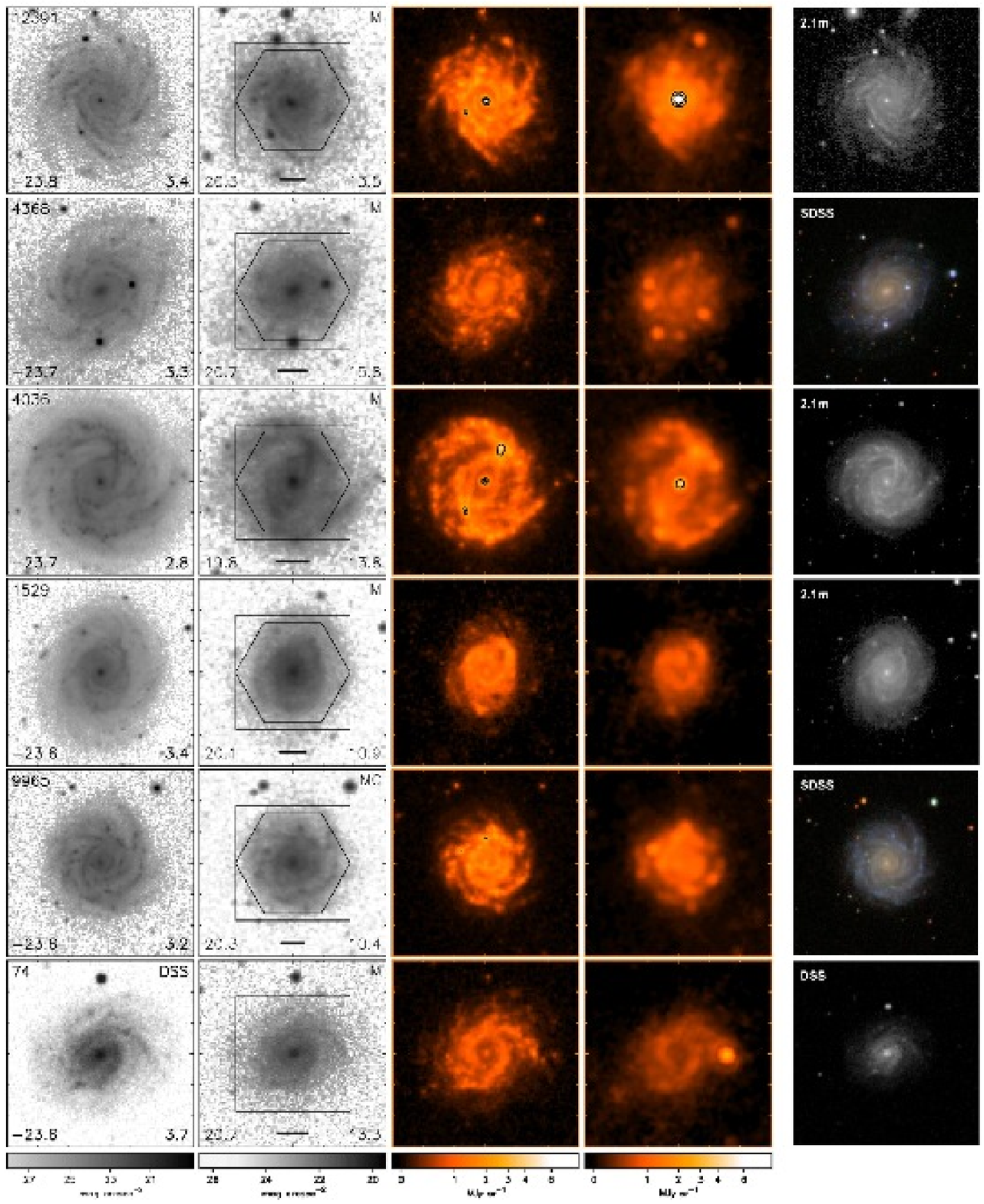}
\caption{continued [COLOR]}
\end{figure}

\clearpage

\begin{figure}
\figurenum{9}
\epsscale{1.0}
\plotone{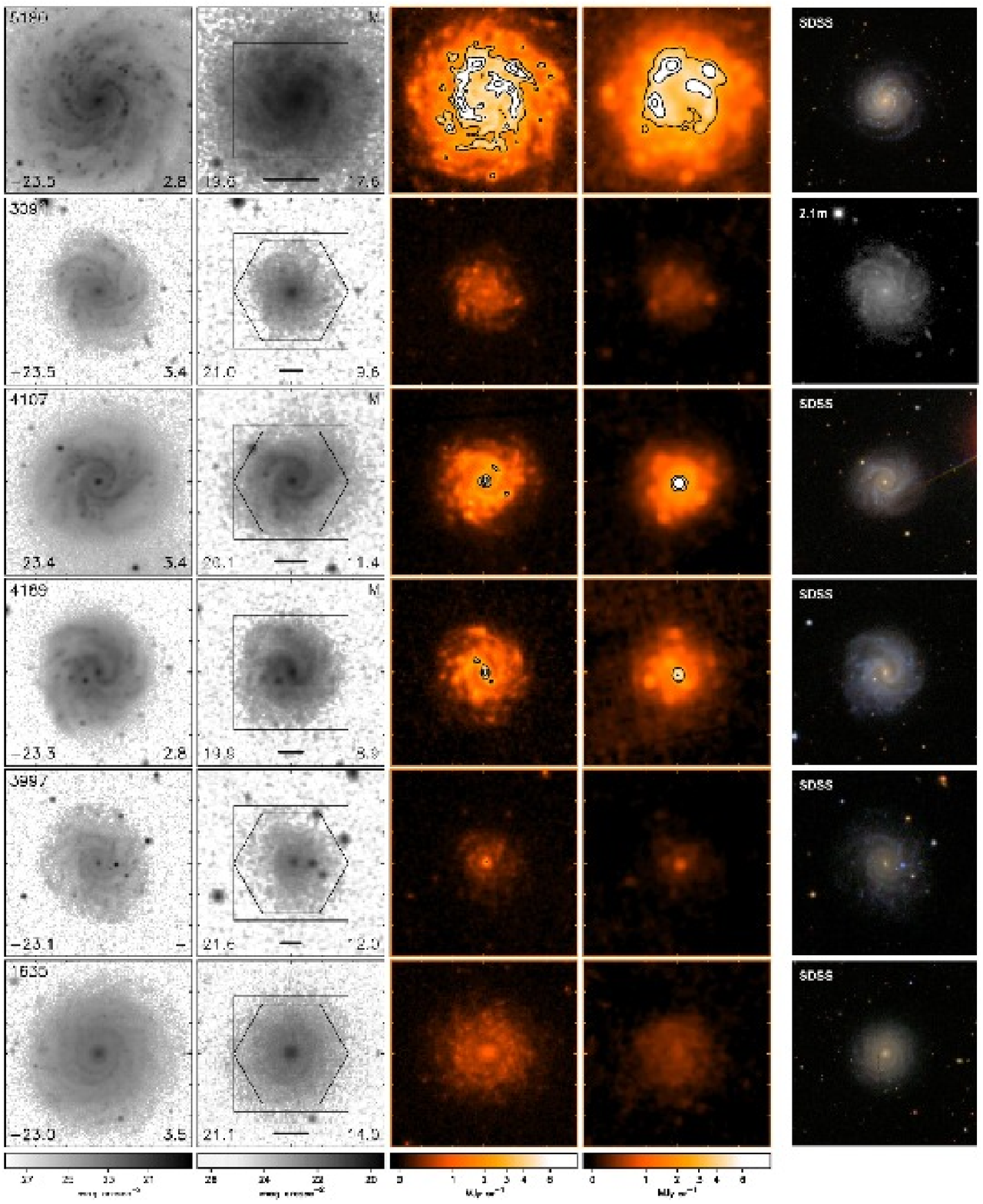}
\caption{continued [COLOR]}
\end{figure}

\clearpage

\begin{figure}
\figurenum{9}
\epsscale{1.0}
\plotone{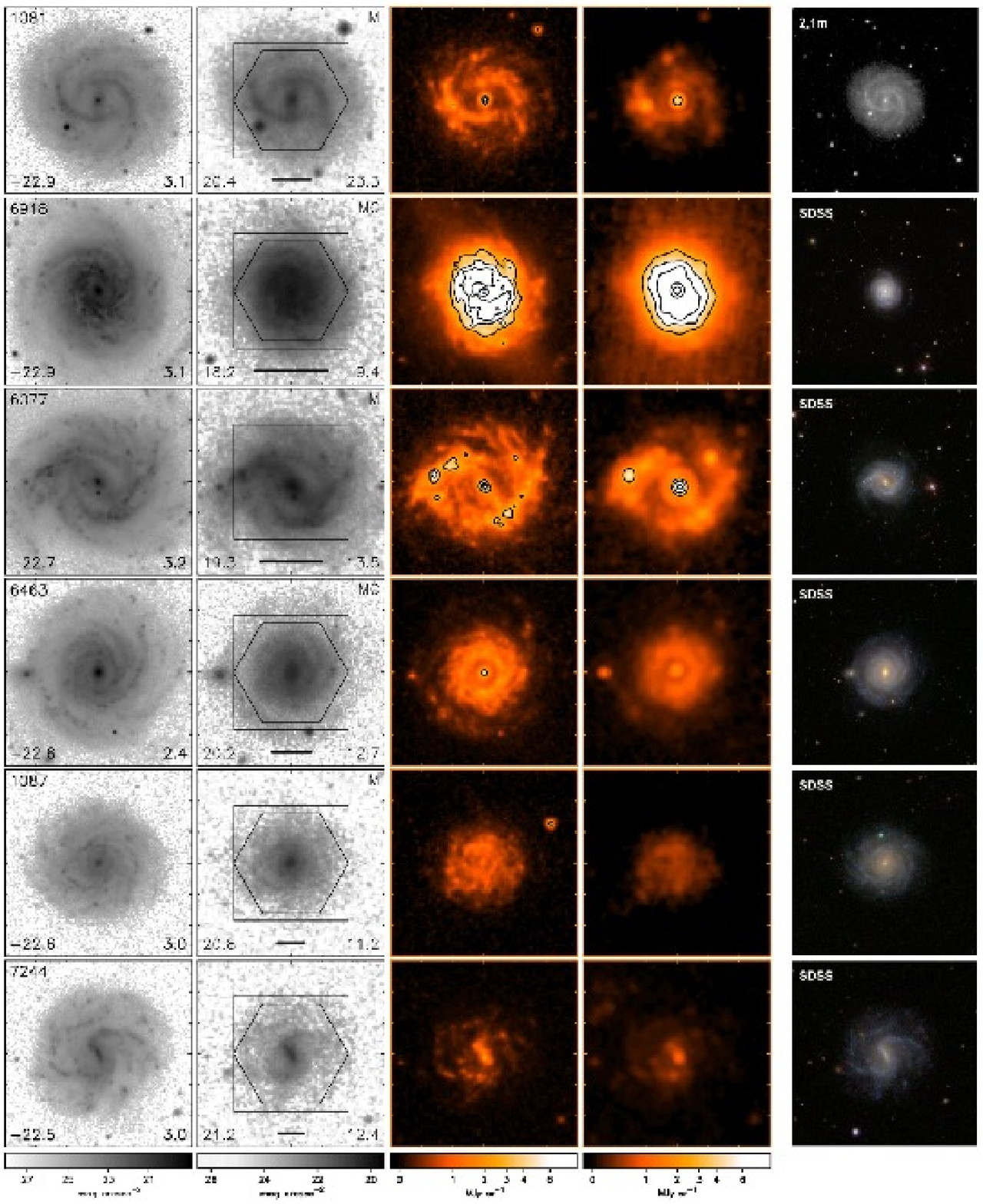}
\caption{continued [COLOR]}
\end{figure}

\clearpage

\begin{figure}
\figurenum{9}
\epsscale{1.0}
\plotone{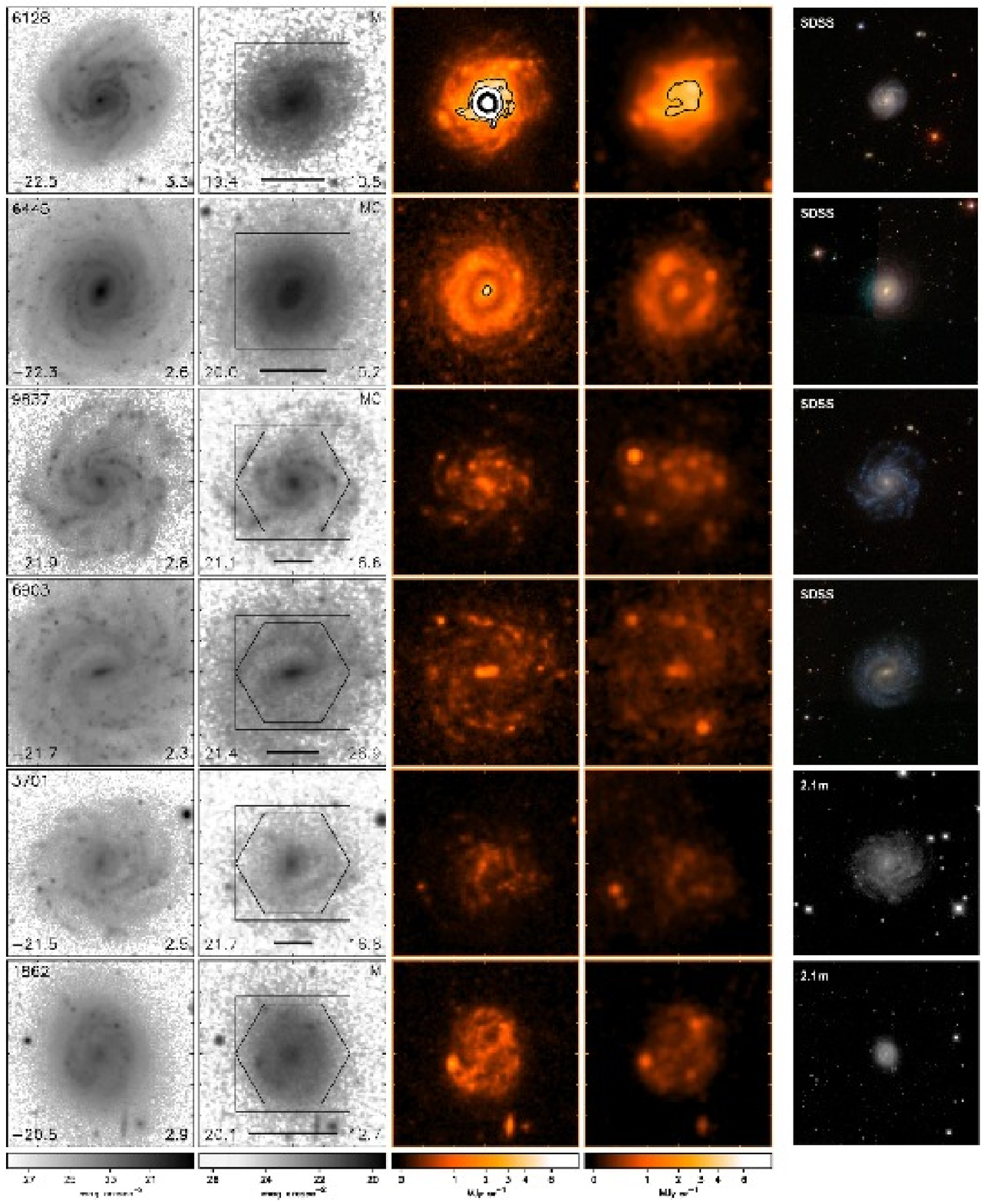}
\caption{continued [COLOR]}
\end{figure}

\clearpage

\begin{figure}
\figurenum{10}
\epsscale{1}
\plotone{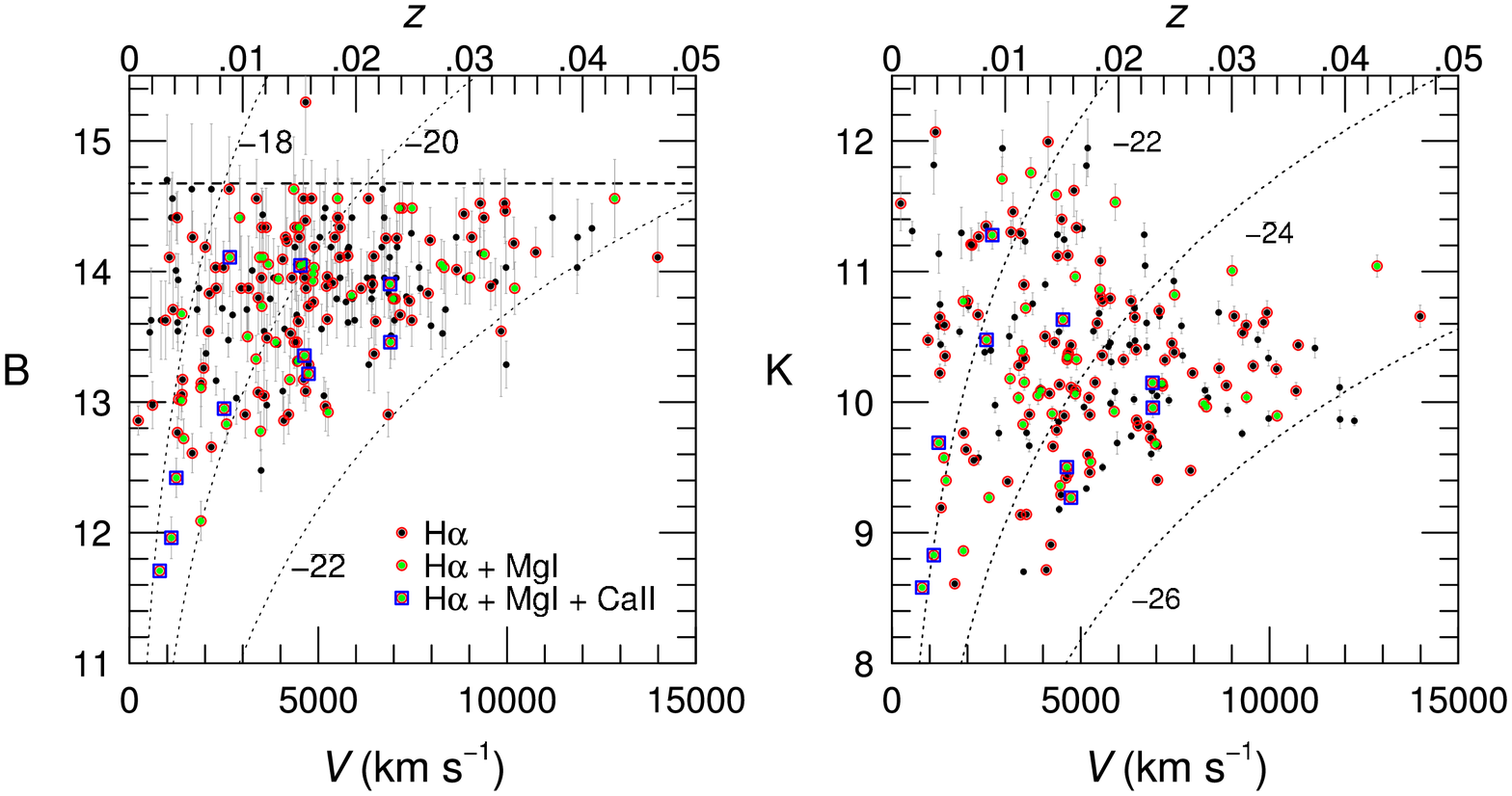}
\caption{Hubble diagrams for the DMS in the $B$-band (left) and
  $K$-band (right). Sub-samples are marked according to the key: black
  points (Phase-A sample); red-encircled points (H$\alpha$ sample);
  green-cored points (Phase-B \ion{Mg}{1b} sample); blue-ensquared
  points (Phase-B \ion{Ca}{2} sample). Objects in the \ion{Mg}{1b}
  sub-sample are also in the H$\alpha$ sub-sample; objects in the
  \ion{Ca}{2} sub-sample are also in \ion{Mg}{1b} and H$\alpha$
  sub-samples. Dotted lines of constant luminosity are marked. The
  $B$-band completeness limit of the UGC is delineated with a
  horizontal dashed line in the left panel. [COLOR]}
\end{figure}

\clearpage

\begin{figure}
\figurenum{11}
\epsscale{0.7}
\plotone{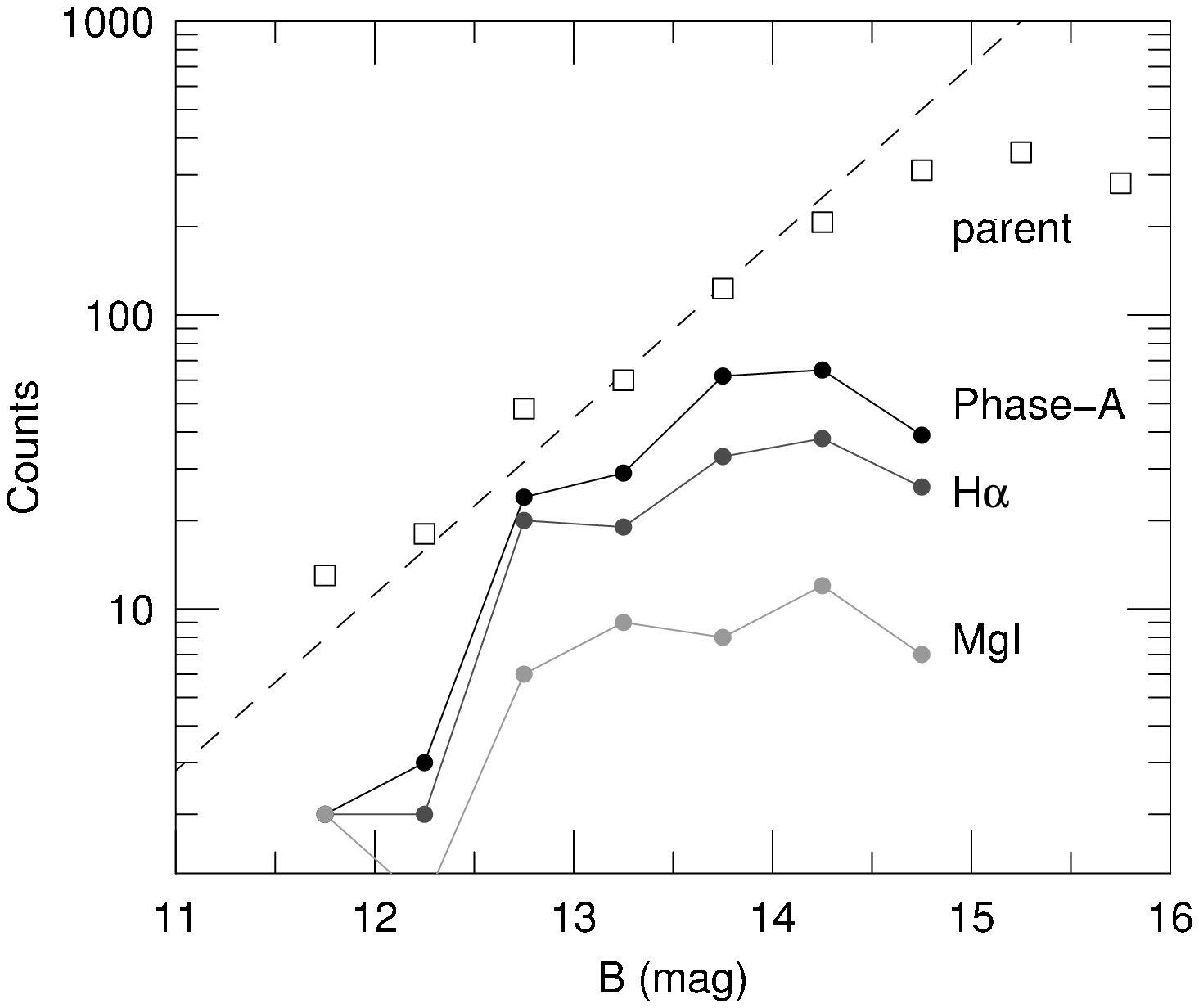}
\caption{$B$-band galaxy counts (per 0.5 mag) for sub-samples in our
  survey selected from the UGC. The dashed line has a Euclidean slope
  of 0.6 dex, with an arbitrary normalization. The curved marked
  ``H$\alpha$'' represents the Phase-A sample with H$\alpha$ IFS.  The
  curve marked ``MgI'' is the subset of the Phase-B sample with
  \ion{Mg}{1} IFS.}
\end{figure}

\clearpage

\begin{figure}
\figurenum{12}
\epsscale{1.0}
\plotone{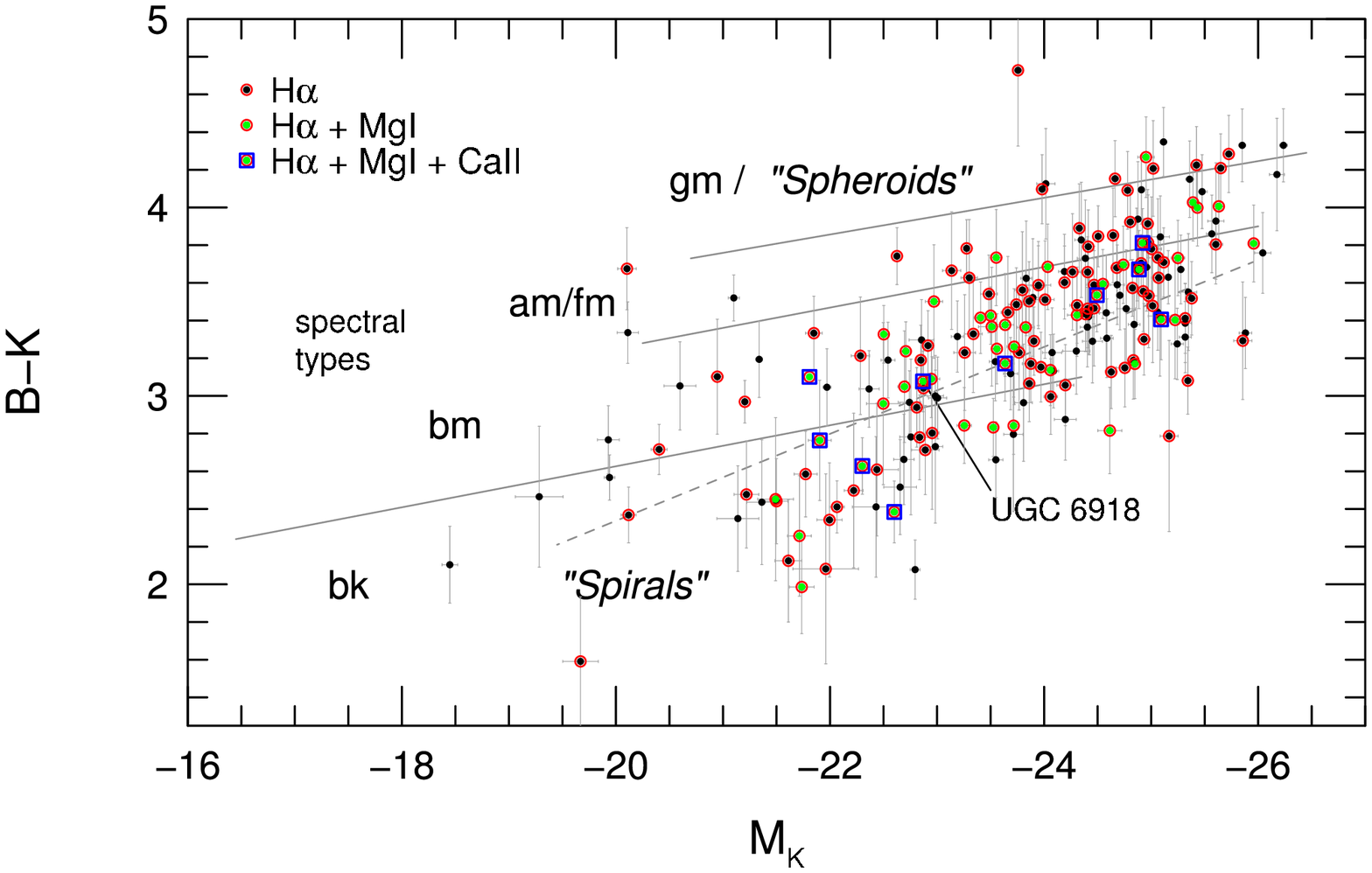}
\caption{Optical--near-infrared color--absolute-magnitude diagram for
  the DMS, with sources marked as in Figure 10. Trends for spirals,
  spheroidals, and galaxies classified by spectral type from Bershady
  (1995) are marked with lines over the observed luminosity range in
  that survey, and labeled for reference (see text). The location of
  UGC 6918 is marked. [COLOR]}
\end{figure}

\clearpage

\begin{figure}
\figurenum{13}
\epsscale{1}
\plotone{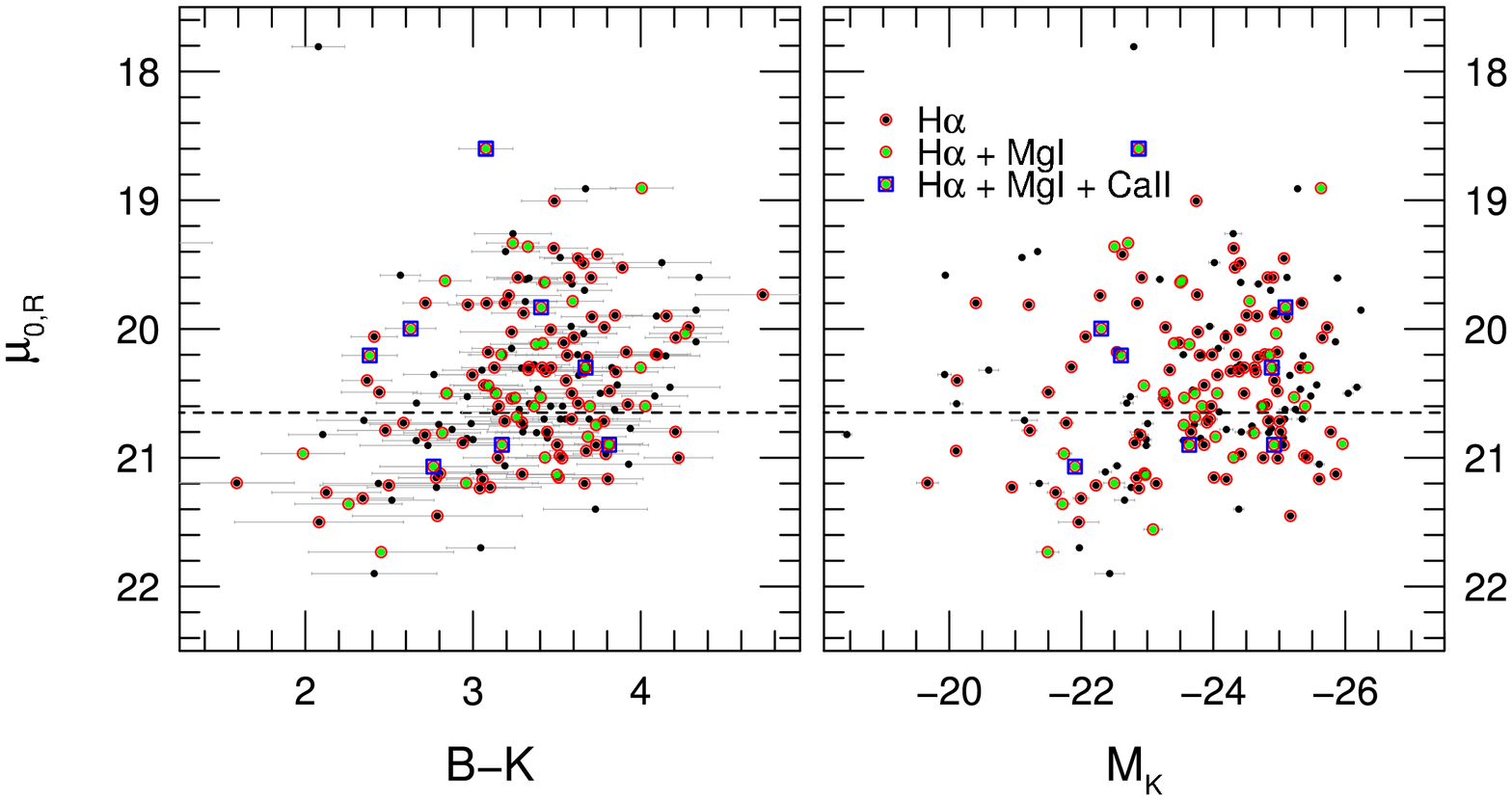}
\caption{Distributions in $B-K$ color, $K$-band absolute magnitude,
  $M_K$, and red disk central surface-brightness, $\mu_{0,R}$ for
  the DMS, with sources marked as in Figure 10. The approximate Freeman
  surface-brightness is indicated by the horizontal dashed line. [COLOR]}
\end{figure}

\clearpage

\begin{figure}
\figurenum{14}
\epsscale{1.0}
\plotone{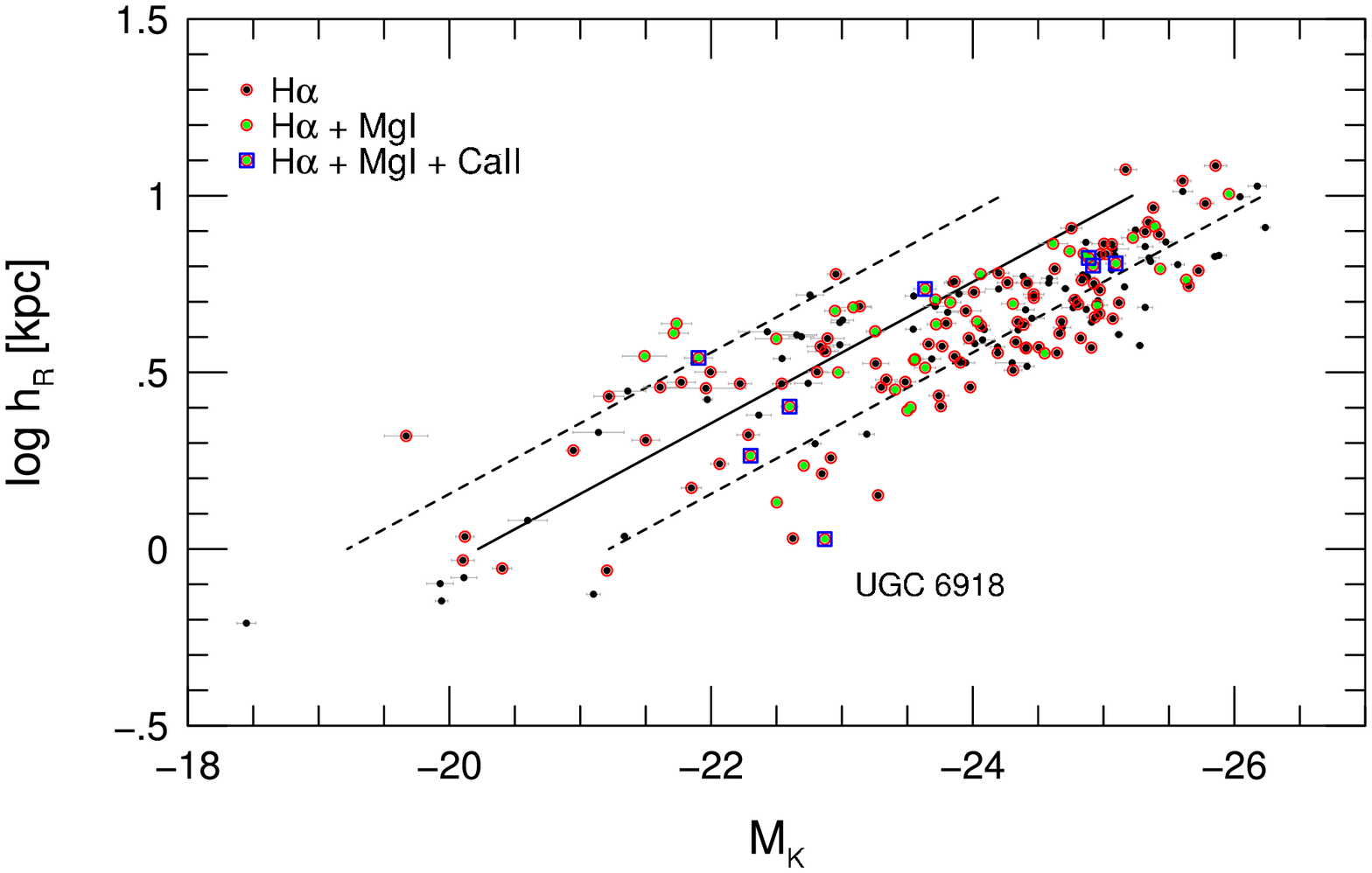}
\caption{Size versus luminosity for the DMS, with sources marked as in
  Figure 10.  Size corresponds to the disk scale-length, $h_R$,
  measured on red POSS images. Luminosity is measured in the
  $K$-band. Diagonal lines correspond to constant mean
  surface-brightness assuming disks dominate the integrated luminosity
  for Freeman surface-brightness disks of color $B-K = 2.3$, 3.3, and
  4.3 (left to right, respectively). [COLOR]}
\end{figure}

\clearpage

\begin{figure}
\figurenum{15}
\epsscale{1.0}
\rotate
\plotone{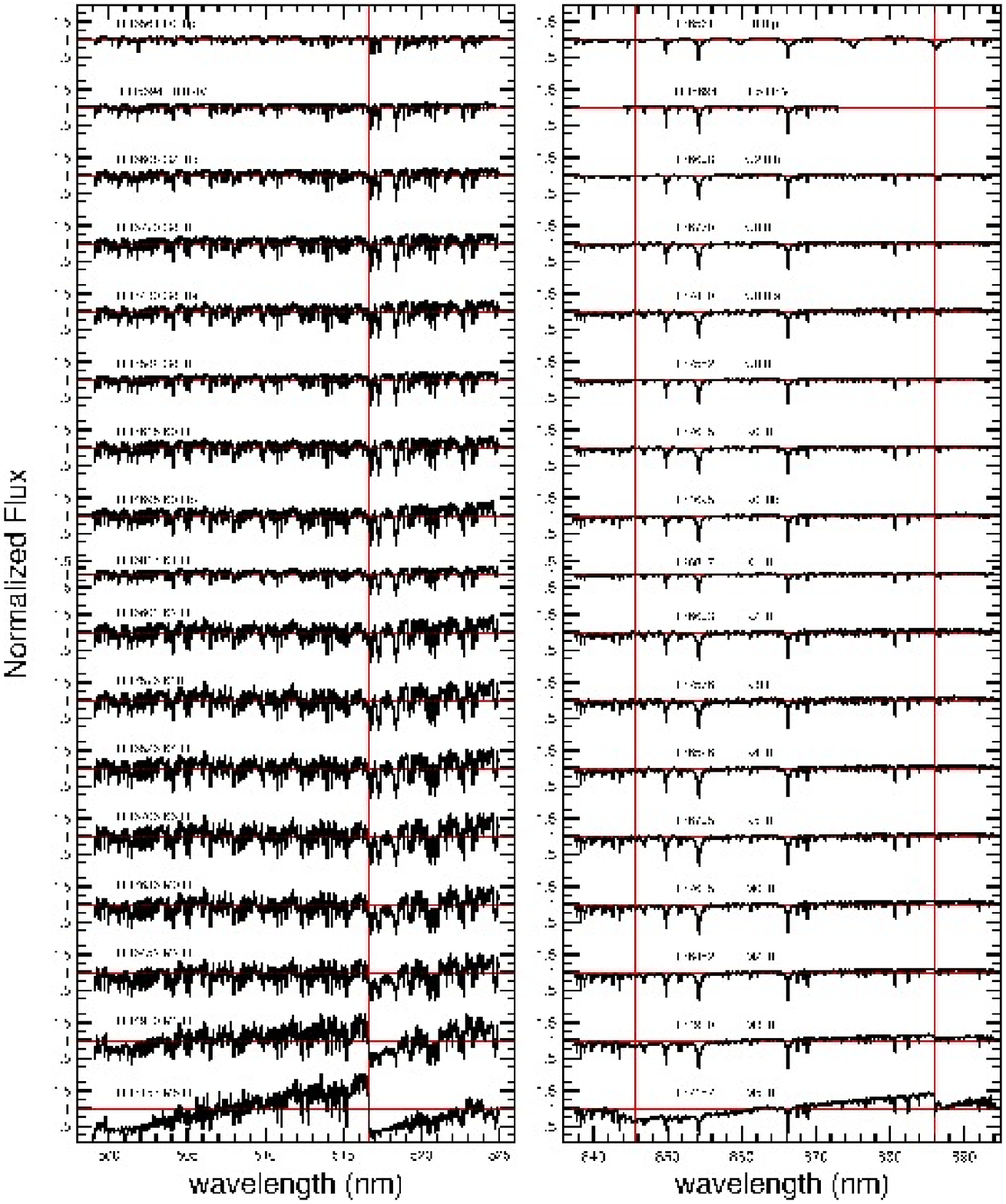}
\caption{SparsePak template subset for luminosity-class II-III and
  spectral types F0 to M5. Spectra are normalized to unity mean.
  Vertical lines mark molecular band-heads. [COLOR IN ELECTRONIC
    EDITION ONLY.]}
\end{figure}

\clearpage

\begin{figure}
\figurenum{16}
\epsscale{0.8}
\plotone{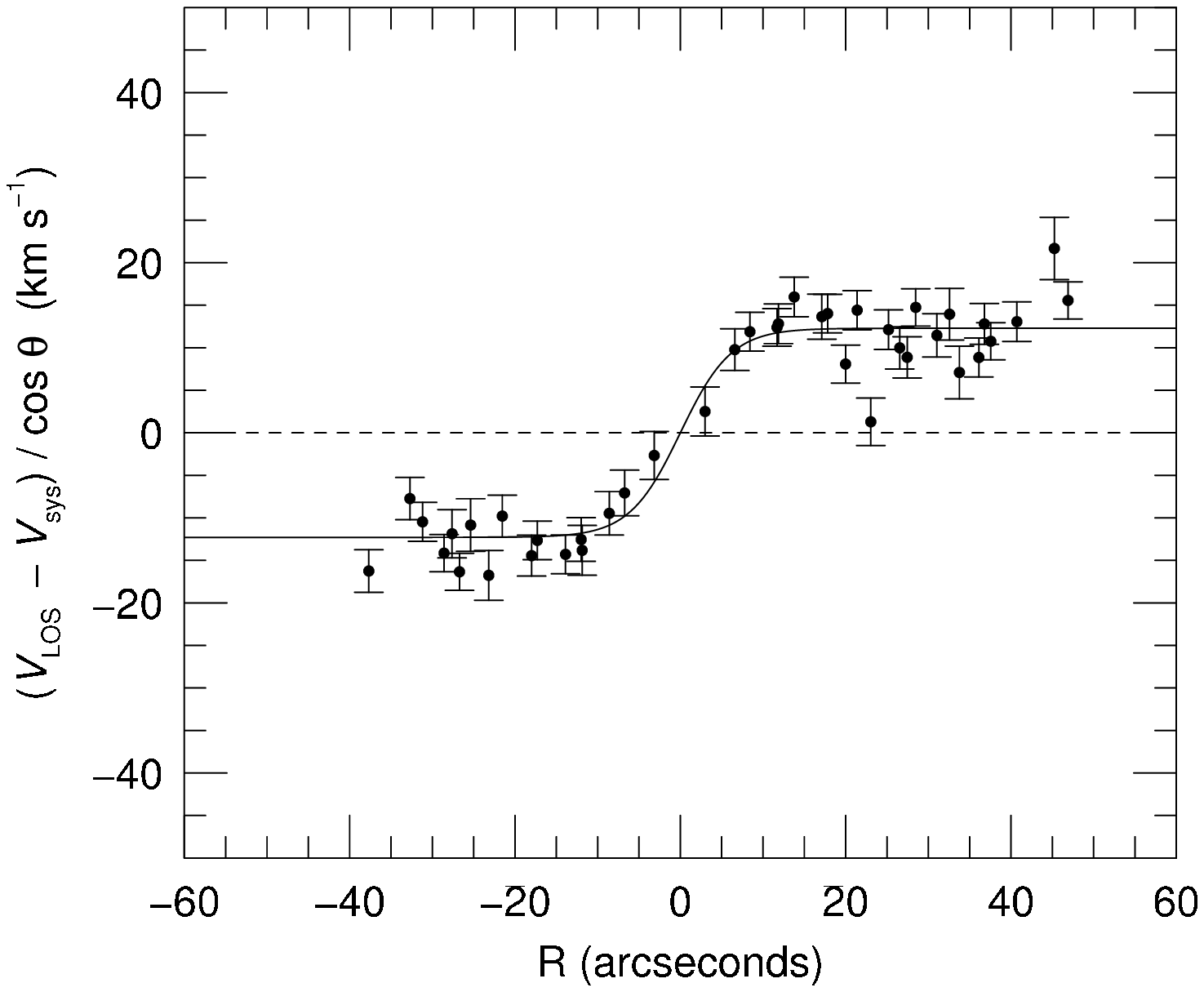}
\caption{H$\alpha$ {\it projected} rotation curve for UGC 6897 as an
  example of the very low-levels of projected rotation that can be
  detected using SparsePak and the echelle grating, and used to derive
  precise inclinations from iTF. Line-of-sight velocities ($\vlos$)
  are extracted from $\pm 30^\circ$ wedge about the determined
  kinematic major axis, corrected for systemic recession velocity
  ($\vsys$), and corrected for azimuthal projection (only). A model
  rotation curve with projected amplitude of 12.3$\pm$0.7 km s$^{-1}$
  about the barycenter is shown as a solid line.}
\end{figure}

\clearpage

\begin{figure}
\figurenum{17}
\epsscale{0.6}
\plotone{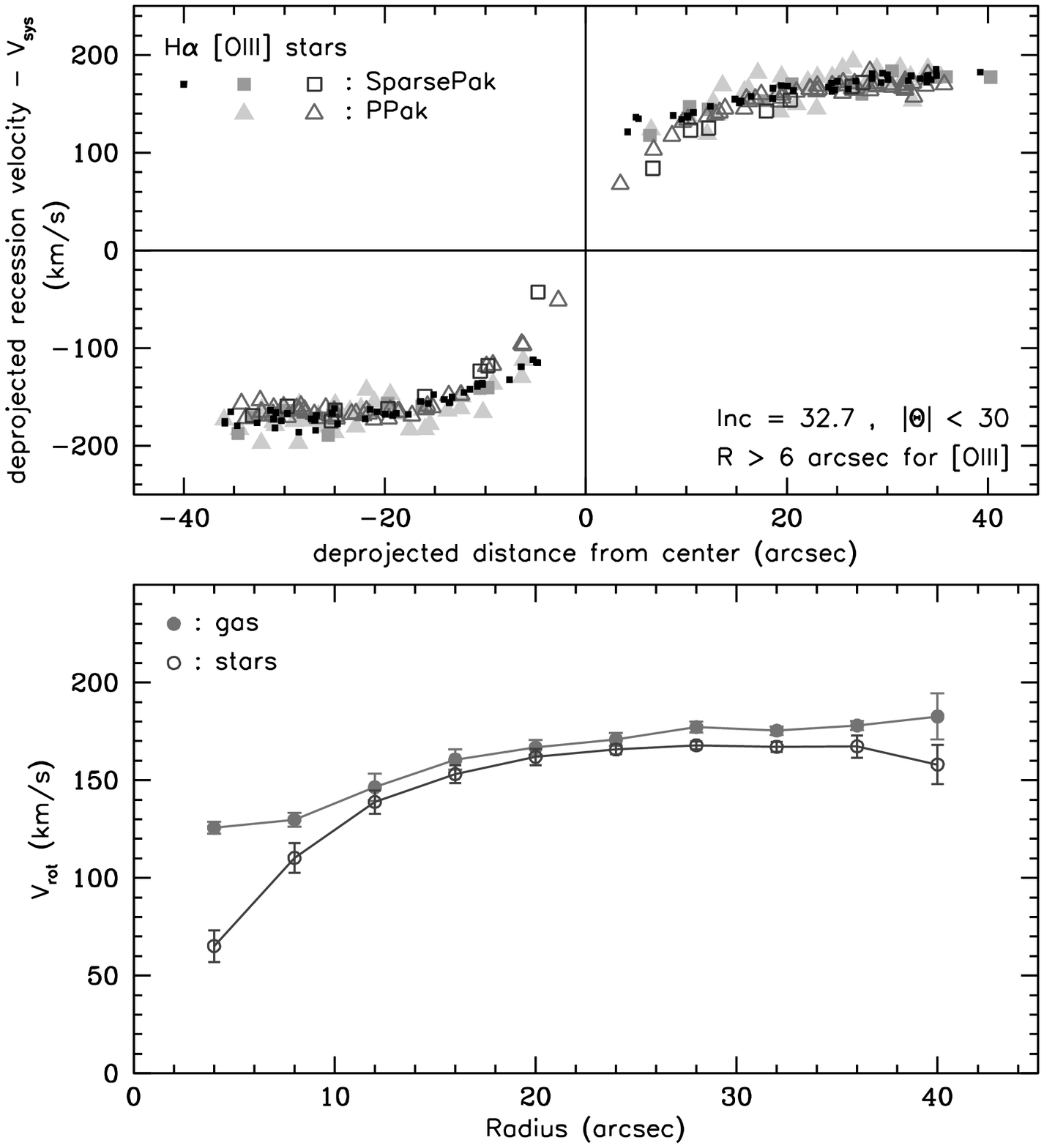}
\vskip 0.75in
\caption{Asymmetric drift between the de-projected tangential
  velocities of stars (open squares and triangles) and ionized gas
  (filled grey squares and triangles for [\ion{O}{3}]$\lambda$5007;
  small, filled squares for H$\alpha$) in UGC 6918, as observed with
  SparsePak and PPak.  Gas velocities are based on line centroids,
  while stellar velocities are from cross-correlation with stellar
  templates in the same (\ion{Mg}{1b}) spectral region. The top panel
  shows deprojected tangential velocities for individual fibers within
  30$^\circ$ of the kinematic major axis, adopting a kinematic
  inclination of 32.7$^\circ$, and corrected for the systemic
  recession velocity, $\vsys$. The bottom panel gives the amplitude of
  the deprojected tangential velocity for gas and stars, $\vrot$,
  derived from tilted ring fits to radial bins of fibers, adopting the
  same inclination for each bin, and combining all SparsePak and PPak
  data shown in the top panel.}
\end{figure}

\clearpage


\begin{figure}
\figurenum{18}
\epsscale{1}
\plotone{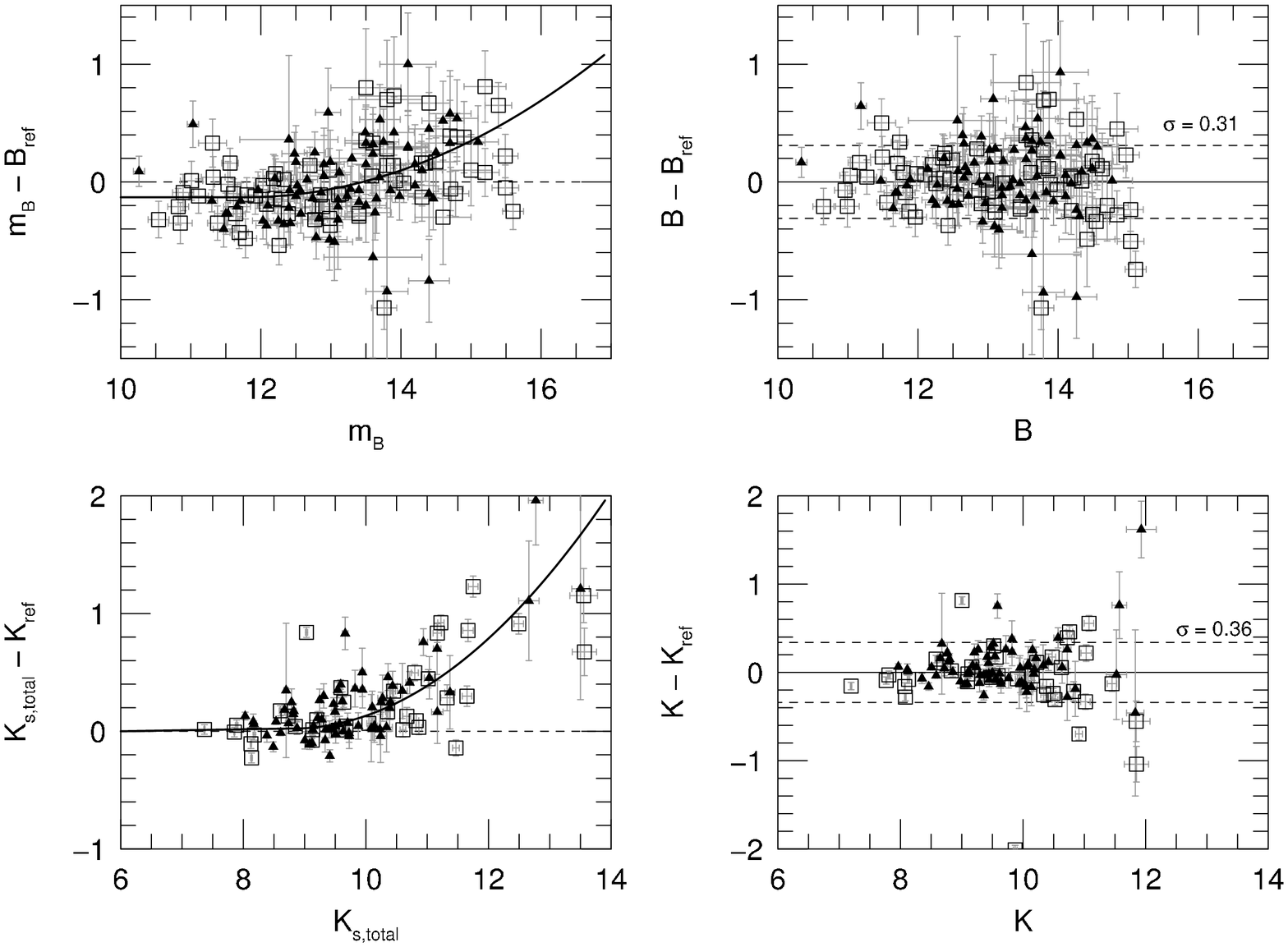}
\caption{Calibration data for RC3 $B$ and 2MASS $K$ total galaxy
  magnitudes.  Left-hand panels show RC3 $m_B$ vs. reference $B$-band
  photometry (top) and 2MASS $K_{\rm s,total}$ vs. reference $K$-band
  photometry (bottom). Reference data are from deep surface-photometry
  of Verheijen \& Sancisi (2001; open squares) and de Jong (1994;
  filled circles). Quadratic fits to deviations are shown as solid
  lines.  Right-hand panels show corrected $B-$ and $K$-band photometry
  vs. the reference photometry, with dashed lines giving the rms
  scatter.}
\end{figure}


\begin{references}

\reference{} Aaronson, M. 1978, ApJ, 221, L103

\reference{} Andersen, D. R. 2001, Ph.D. Thesis, Penn State University

\reference{} Andersen, D. R., Bershady, M. A., Sparke, L. A.,
Gallagher, J. S., III, Wilcots, E. M. 2001, ApJ, 551, L131

\reference{} Andersen, D. R., \& Bershady, M. A. 2003, ApJ, 599, L79

\reference{} Andersen, D. R., Bershady, M. A., Sparke, L. S., Gallagher, J. S., Wilcots, E. M.,
van Driel, W., Monnier-Ragaigne, D. 2006, ApJS, 166,505 

\reference{} Andersen, D. R. \& Bershady, M. A. 2009, ApJ, 700, 1626

\reference{} Baggett, W. E., Baggett, S. M., Anderson, K. S. J. 1998, 116, 1626

\reference{} Bahcall, J, \& Casertano, S. 1984, ApJ, 284, L35

\reference{} Barth, A., Ho, L. C., Sargent, W. L. W. 2002, AJ, 124, 2607

\reference{} Begeman, K. G. 1989, A\&A, 223, 47

\reference{} Bell, E. F., \& De Jong, R. S. 2001, ApJ, 550, 212 

\reference{} Bell, E. F.,  McIntosh, D. H., Katz, N., Weinberg, M. D.
2003, ApJS, 149, 289

\reference{} Bershady, M. A., Hereld, M., Kron, R. G., Koo, D. C.,
Munn, J. A., Majewski, S. R. 1994, AJ, 108, 870

\reference{} Bershady, M. A. 1995, AJ, 109, 87

\reference{} Bershady, M. A., Lowenthal, J. D., Koo, D. C. 1998, ApJ, 505, 50

\reference{} Bershady, M. A., Jangren, A., Conselice, C. J. 2000, AJ,
119, 2645

\reference{} Bershady, M. A., Verheijen, M. A. W., Andersen, D. R., 2002, ASPC, 275, 43

\reference{} Bershady, M. A., Andersen, D. R., Harker, J., Ramsey,
L. W., Verheijen, M. A. W. 2004, PASP, 116, 565

\reference{} Bershady, M. A., Andersen, D. R., Verheijen, M. A. W.,
Westfall, K. B., Crawford, S. M., Swaters, R. A. 2005, ApJS, 156, 311

\reference{} Bershady, M. A., \& Andersen, D. R. in ``Island
Universes,'' ed. R. de Jong, (Springer), 353

\reference{} Bottema, R. 1989, A\&A, 221, 236

\reference{} Bottema, R. 1993, A\&A, 275, 16

\reference{} Bottema, R. 1997, A\&A, 328, 517

\reference{} Braine, J. et al. 1993, A\&AS, 97, 887

\reference{} Bruzual, G., 2007, in ASP Conf. Ser. 374, From Stars to
Galaxies: Building the Pieces to Build Up the Universe, ed.
A. Vallenari, R. Tantalo, L. Portinari \& A. Moretti (san Francisco:
ASP), 303

\reference{} Bruzual, G., Charlot S. 2003, MNRAS, 344, 1000

\reference{} Casoli, E., et al. 1998, A\&A, 331, 451

\reference{} Courteau, S., \& Rix, H.-W. 1999, ApJ, 513, 561

\reference{} Conroy, C., Gunn, J. E., White, M. 2009, ApJ, 699, 486

\reference{} Courteau, S., Andersen, D. R., Bershady, M. A.,
MacArthur, L. A., Rix, H.-W. 2003, ApJ, 594, 208

\reference{} de Blok, W. J. G., Walter, F., Brinks, E., Trachternach,
C., Oh, S.-H., Kennicutt, R. C. 2008, AJ, 136, 2648

\reference{} de Grijs, R., van der Kruit, P. C. 1996, A\&AS, 117, 19

\reference{} de Grijs, R., van der Kruit, P. C. 1997, A\&AS, 327, 966

\reference{} de Jong, R. 1994, A\&AS, 106, 451

\reference{} de Vaucouleurs, G., de Vaucouleurs, A., Corwin, H., Buta,
R., Paturel, G. \& Foqu\'{e}, P. 1991, Third Reference Catalogue of
Bright Galaxies (Springer, New York)

\reference{} de Bruyne, V., De Rijcke, S., Dejonghe, H., Zeilinger,
W. W. 2004, MNRAS, 349, 461

\reference{} Freeman, K. C. 1970, ApJ, 160, 811

\reference{} Graham, A. 2001, ApJ, 121, 820

\reference{} Herrmann, K. A., and Ciardullo, R. 2009, ApJ, accepted
(arXiv:0910.0266v1)

\reference{} Jansen, R. A., Franx, M., Fabricant, D., Caldwell,
M. 2000, ApJS, 126, 271

\reference{} Kelz, A. et al. 2006, PASP, 118, 129

\reference{} Kennicutt et al. 2003, PASP, 115, 928

\reference{} Kranz, T., Slyz, A., Rix, H.-W. 2001, ApJ, 562, 164

\reference{} Kregel, M., van der Kruit, P. C., de Grijs, R. 2002,
MNRAS, 334, 646

\reference{} Larson, R. B., \& Tinsley, B. M. 1978, ApJ, 219, 46

\reference{} Le Borgne, D., Rocca-Volmerange, B., Prugniel, P.,
Lan\c{c}on, A., Fioc, M., and Soubiran, C. 2004, A\&A, 425, 881

\reference{} Maller, A., Simard, L., Guhathakurta P., Hjorth, J.,
Jaunsen, A. O., Flores, R., Primarck, J. R. 2000, ApJ, 533, 194

\reference{} Maraston, C. 2005, MNRAS, 362, 799

\reference{} McGaugh, S. S. 2005, ApJ, 632, 859 

\reference{} Navarro, J. F., Frenk, C. S., White, S. D. M. 1997, ApJ, 490, 493

\reference{} Nilson, P. 1973, Uppsala General Catalogue of Galaxies,
Uppsala Astr. Obs. Ann., Vol. 6 (UGC)

\reference{} Noordermeer, E., Verheijen, M. A. W. 2007, MNRAS, 381, 1463

\reference{} Pfenniger, D., Combes, F. 1994, A\&A, 285, 94

\reference{} Pohlen, M. and Trujillo, I. 2005, A\&A, 454, 759

\reference{} Portinari, L., Sommer-Larsen, J., Tantalo, R. 2004, MNRAS, 347, 691

\reference{} Rix, H.-W., Zaritsky, D. 1995, ApJ, 447, 82

\reference{} Sanders, R. H., \& Verheijen, M. A. W. 1998, ApJ, 503, 97

\reference{} Schlegel, D. J., Finkbeiner, D. P., Davis, M. 1998, ApJ, 500, 525

\reference{} Skrutskie, M. F. et al. 2006, AJ, 131, 1163

\reference{} Spekkens, K., \& Sellwood, J. A. 2007, ApJ, 664, 204

\reference{} Swaters, R. A. Schoenmakers, R. H. M., Sancisi, R., van
Albada, T. S. 1999, MNRAS, 304, 330

\reference{} Swaters, R. A., \& Balcells, M. 2002, A\&A, 390, 829


\reference{} Trachternach, C., de Blok, W. J. G., Walter, F., Brinks, E., Trachternach,
C., Oh, S.-H., Kennicutt, R. C. 2008, AJ, 136, 2720

\reference{} Tully, R. B., \& Fisher, J. R. 1977, A\&A, 54, 661

\reference{} van Albada, T. S., Bahcall, J. N., Begeman, K., Sancisi,
R. 1985, ApJ, 295, 305

\reference{} van Albada, T. S., Sancisi, R., Petrou, M. Tayler,
R. J. 1986, Royal Society (London), Philos. Trans., Series A, 320, \#
1556, 447

\reference{} van der Kruit, P. C., \& Searle, L. 1981, A\&A, 95, 105

\reference{} van der Kruit, P. C., \& Freeman, K. C. 1984, ApJ, 278, 81

\reference{} van der Kruit, P. C., \& Freeman, K. C. 1986, ApJ, 303, 556

\reference{} Verheijen, M. A. W. 1997, Ph.D. thesis, University of Groningen

\reference{} Verheijen, M. A. W. 2001, ApJ, 563, 694

\reference{} Verheijen, M. A. W., \& Sancisi, R. 2001, A\&A, 370, 765

\reference{} Verheijen, M. A. W., Bershady, M. A., Andersen,
D. R. 2003, in ``The Mass of Galaxies at Low and High Redshift,''
eds. R. Bender and A. Renzini (Springer-Verlag), 221

\reference{} Verheijen, M. A. W., Bershady, M. A., Andersen, D. R.,
Swaters, R. A., Westfall, K., Kelz, A., Roth, M. M. 2004, AN, 325

\reference{} Weiner, B. J., Sellwood, J. A., Williams, T. B. 2001, ApJ, 546, 931

\reference{} Westfall, K. B. 2009, Ph.D. Thesis, Unversity of Wisconsin

\reference{} York, D. G. et al. 2000, AJ, 120, 1579

\reference{} Zibetti, S., Charlot, S., Rix, H.-W. 2009, submitted to
MNRAS, astro-ph/arXiv:0904.4252v1

\reference{} Zwaan, M. A., van der Hulst, J. M., de Blok, W. J. G.,
McGaugh, S. S. 1995, MNRAS, 273, L35

\end{references}
\end{document}